\documentclass[preprint,12pt]{elsarticle}

\usepackage{amssymb}
\usepackage{lineno}
\usepackage{tablefootnote}
\usepackage{mathtools}
\usepackage[colorlinks]{hyperref}
\usepackage[colorinlistoftodos]{todonotes}
\usepackage{verbatim}
\usepackage{longtable}
\usepackage{caption}
\usepackage{svg}
\usepackage{algorithm}
\usepackage{algpseudocode}

\usepackage{xcolor, color, soul}
\sethlcolor{yellow}
\journal{Energy and AI}

\begin{document}

\begin{frontmatter}



\title{Fuelprop: Fuel property prediction from ATR-FTIR spectroscopic data}


\author[inst1]{Mohammed Almomtan}
\ead{mohammed.almomtan@kaust.edu.sa}
\affiliation[inst1]{organization={Clean Energy Research Platform, Physical Sciences and Engineering (PSE) Division}, 
            addressline={King Abdullah University of Science and Technology (KAUST)}, 
            city={Thuwal},
            postcode={23955-6900}, 
            country={Saudi Arabia}}

\author[inst1]{Emad Al Ibrahim}
\ead{emad.ibrahim@kaust.edu.sa}
\author[inst1]{Aamir Farooq\corref{cor1}}
\ead{aamir.farooq@kaust.edu.sa}
\cortext[cor1]{Corresponding author}

\begin{abstract}
Synthetic fuels are crucial for decarbonizing the transportation sector. A significant challenge lies in the rapid and efficient characterization of these fuels. Chemometric methods using ATR-FTIR data offer a potential alternative to conventional techniques. This study expands the applicability and performance of chemometric models by providing an extensive ATR-FTIR spectral dataset and exploring various data enhancement strategies. Data enhancement was achieved by semi-supervised data generation, consistency enforcement through unsupervised data augmentation, and data imputation using synthetic spectra blending and pseudo-labeling. Models were trained on surrogate fuels and rigorously tested on real fuels, representing out-of-distribution testing conditions. We believe that this work will enhance the adoption of chemometric models for fuel characterization.

\end{abstract}

\begin{graphicalabstract}
\includegraphics[width=1 \textwidth]{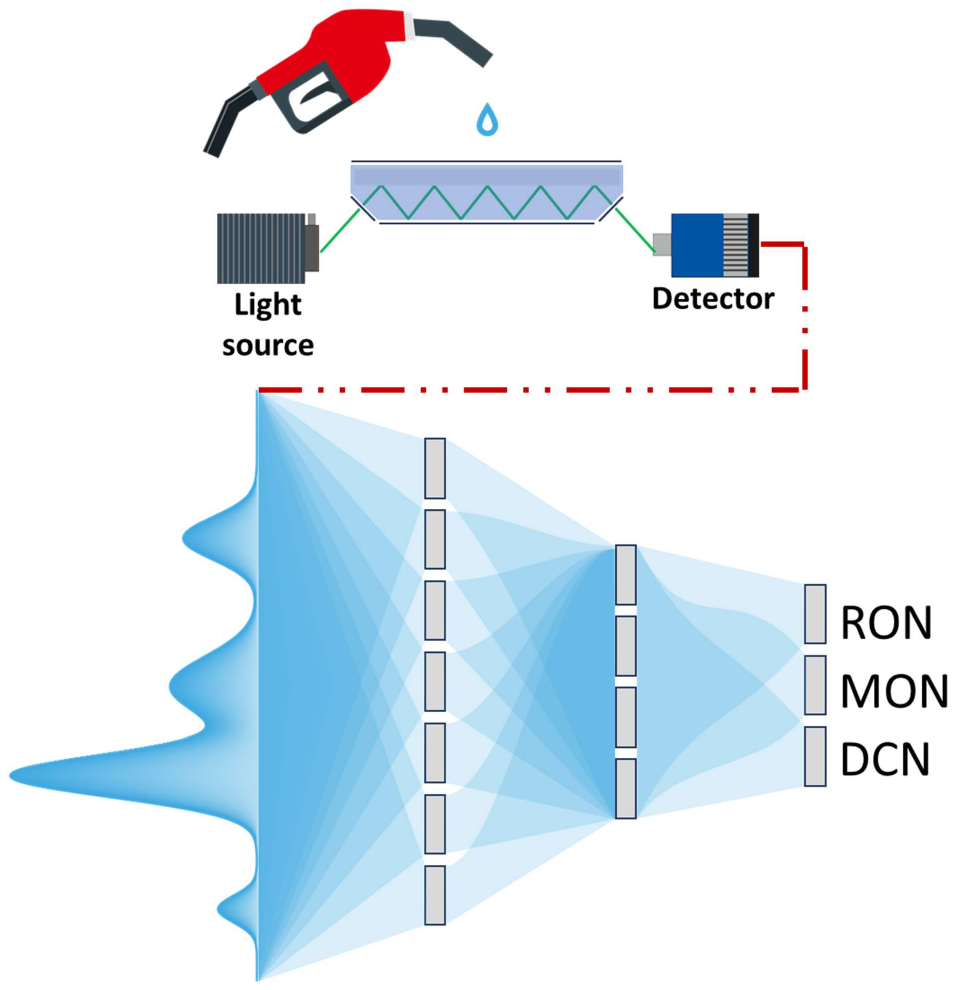}
\end{graphicalabstract}

\begin{highlights}
\item ATR-FTIR spectral measurements for pure components, surrogate blends, and real fuels are reported.
\item A CNN model is trained on surrogates and evaluated on real fuels for property prediction.
\item Synthetic spectra blending is used to impute missing spectral measurements.
\item Pseudo-labeling is applied to expand the training data and improve accuracy.
\item Consistency enforcement through unsupervised data augmentation is employed to enhance model generalization to real fuels.
\end{highlights}


\end{frontmatter}


\textbf{Abbreviations}
Research Octane Number (RON); Motor Octane Number (MON); Derived Cetane Number (DCN); Internal Combustion Engine (ICE); Heavy-Duty Vehicle (HDV); Greenhouse Gas (GHG); Cooperative Fuel Research (CFR); Primary Reference Fuel (PRF); Ignition Quality Tester (IQT); Constant Volume Combustion Chamber (CVCC); Fourier Transform Infrared (FTIR); Near-Infrared (NIR); Attenuated Total Reflection (ATR); Partial Least Squares Regression (PLSR); Artificial Neural Network (ANN); Principal Component Analysis (PCA); Convolutional Neural Network (CNN); Out-of-Distribution (OOD); Mean Absolute Error (MAE); Interquartile Range (IQR); Hydrogen ($H_2$); Carbon Dioxide ($CO_2$); paraffinic, iso-paraffinic, olefinic, naphthenic, aromatic, and oxygenate (PIONA-O).
\section{Introduction}

Fossil fuels have been the primary source of energy for transportation, enabling the mobility of people and goods around the world. In fact, oil currently accounts for $92\%$ of all road transport energy demand and $85\%$ of non-road transport energy demand \cite{IEA}. While combustion of these fuels powers transportation, it also releases harmful pollutants into the environment, contributing to climate change and air pollution. More than $20\%$ of all global $CO_2$ emissions come from transportation-related fossil fuel consumption \cite{EDGAR}. This clash between mobility and the environment has become a crucial issue in today's society. Finding ways to reduce the negative impact of transportation on the environment while still meeting mobility needs is a challenge that requires innovative solutions. Current policy developments and regulatory standards are driving the accelerated electrification of light-duty vehicle (LDV) fleet \cite{EIA, raimi2024global}. However, projections estimate that the fleet of internal combustion engine (ICE) vehicles will retain no less than $70-80\%$ of the LDV market share by 2035 even with an aggressive transition towards electrification \cite{bp, EVprojection}. The heavy-duty vehicle (HDV) fleet which includes activities like shipping, long-haul trucking, and aviation is even less likely to be electrified with projections indicating no significant electrification of the HDV fleet until 2050 \cite{ExxonMobil}. In fact, for the near future at least, the decline in oil demand and the subsequent decline in greenhouse gas (GHG) emissions in the transportation sector are more likely to be driven by improvements in efficiency and emissions reductions of the ICE vehicle fleet \cite{bp}.

One promising solution to the challenge of reducing the negative impact of ICE-dominated transportation on the environment is the development of synthetic fuels that can substitute fossil fuels. Synthetic fuels are created from non-petroleum sources \cite{ram2023overview}; for example, by combining H2, generated from solar-driven water electrolysis, and CO2, captured from air or power plants. Produced this way, synthetic fuels may be classified as having net-zero carbon emissions. They can also have the potential to burn cleaner and emit fewer pollutants than traditional fossil fuels. To directly replace fossil fuels without additional costs of vehicle modification or infrastructure development, synthetic fuels are tailored to specific applications such as aviation or heavy-duty trucking. The suitability of a candidate synthetic fuel for a specific application can be assessed by evaluating its chemical and physical properties and their alignment with those of current fuels.

Ramirez et al. identified the most important properties that govern engine operation and/or indicate compatibility with existing infrastructure \cite{mani_efuels}. These properties include research octane number (RON), motor octane number (MON), and derived cetane number (DCN). RON and MON are measurements of a fuel's ability to resist knocking in an engine which is relevant for the performance of spark ignition (SI) engines. DCN, on the other hand, is a measurement of a fuel's ignition quality in a compression ignition engine, such as a diesel engine. These measurements are standardized by the American Society for Testing and Materials (ASTM) as ASTM D2699 for RON, ASTM D2700 for MON, and ASTM D6890/D7170/D7668 for DCN \cite{ASTM-RON, ASTM-MON, ASTMDCN, ASTMDCN2, ASTMD7668}. 

The octane number is typically determined using a standardized Cooperative Fuel Research (CFR) engine, operated as a single-cylinder, four-stroke cycle engine with a variable compression ratio. The tendency of the test fuel to knock is evaluated in comparison to a primary reference fuel (PRF) blend, consisting of iso-octane (2,2,4-trimethylpentane) and n-heptane. When the CFR engine operates under standard conditions outlined by ASTM D2699, it measures RON \cite{ASTM-RON}. In contrast, running the engine with preheated fuel at elevated speeds, as defined by ASTM D2700, yields MON \cite{ASTM-MON}.

The derived cetane number (DCN) is determined using a constant volume combustion chamber, commonly referred to as the ignition quality tester (IQT). It is calculated based on the interval between fuel injection and the point of maximum pressure rise rate, as specified in ASTM D6890-16 \cite{ASTMDCN}. Alternatively, DCN can be assessed using a fuel ignition tester (FIT), in accordance with ASTM D7170 \cite{ASTMDCN2}, or through a constant volume combustion chamber (CVCC) method, as outlined in ASTM D7668 \cite{ASTMD7668}.

While the above measurement standards are extensively utilized within the research community, several inherent limitations are associated with their use. Among these is the high capital cost associated with the equipment along with ongoing expenses related to maintenance and calibration. Additionally, the large size of the equipment, the lengthy testing duration, and the complexities involved in their operation further restrict their practicality for rapid assessments. Further, large volumes (as high as 1 L) are generally necessary for comprehensive experimental evaluations. New proposed synthetic fuel candidates are often synthesized in laboratory settings in small quantities, which poses significant challenges for quality assessment.

An alternative method for characterizing fuels is through the use of vibrational spectroscopy. This technique measures the vibrations of molecules in a fuel sample and provides information about its chemical composition. Vibrational spectroscopy has the potential to be a valuable tool in the analysis and development of synthetic fuels, as it allows for the easy, low-cost, rapid, and accurate characterization of new and complex fuels. Most significantly, vibrational spectroscopy has the advantage that measurements can be performed with minimal sample preparation using small volumes, generally less than 1 mL. Various vibrational spectroscopy methods are utilized for fuel analysis such as Fourier Transform Infrared (FTIR) spectroscopy, Near-Infrared (NIR) spectroscopy, Attenuated Total Reflection (ATR) spectroscopy, and Raman spectroscopy. FTIR and ATR spectroscopy are particularly effective in identifying functional groups and characterizing complex fuel mixtures \cite{coury2009atr, xia2019analysis, pena2014fourier}. On the other hand, NIR spectroscopy excels in assessing bulk properties with high throughput \cite{sutliff2024correlating}. Raman spectroscopy complements these techniques by offering insights into molecular symmetry and interactions, making it suitable for detecting trace components \cite{tuschel2014practical}. 

Coupled with chemometric tools such as partial least squares regression (PLSR) and artificial neural networks (ANNs), various vibrational spectra of fuels can be used to predict different properties (RON, MON, DCN, etc.) \cite{DCNVS,NIR,NIR2,NIR3,NIR4,NIR5,FTIR,MIR,Daly1,DCNNIR}. For example, in a study by Al Ibrahim et al., principal component analysis (PCA), PLSR and ANNs were used to predict the octane numbers of blends of hydrocarbons and distillate FACE (Fuels for Advanced Combustion Engines) gasoline blends utilizing gas-phase infrared (IR) spectroscopy data \cite{OctanePaper}. In another study by Al Ibrahim et al., ANNs were used to predict DCN and C/H ratio of blends of hydrocarbon, gasoline, and diesel fuels using gas-phase IR spectra \cite{CetanePaper}. Okada et al., used a least-squares method on gas-phase IR, liquid-phase IR, Raman and combined (gas-phase IR and Raman) spectra to predict DCN among other properties \cite{Sanders}. Wang et al. used Lasso-regularized linear models to estimate DCN among 15 physical and chemical properties based on mid-IR FTIR absorption spectra of different hydrocarbons \cite{newHanson}. Another study by Boddapati et al. used elastic-net regularized linear models to predict ignition delay time (IDT), net heat of combustion (NHC), and DCN from FTIR spectra \cite{Hanson2022}. Dalmiya et al. utilized Convolutional Neural Networks (CNNs) on liquid phase IR spectra to predict DCN of jet fuels and their functional group surrogates \cite{Lynch}. 

Since transportation fuels exist as liquids under ambient conditions, employing ATR infrared spectroscopy presents a practical approach for the chemometric prediction of fuel properties. ATR spectroscopy streamlines the analytical process by allowing direct measurement of the liquid fuel sample, thereby eliminating the need for complex sample preparation steps such as complete vaporization. This method is particularly beneficial for analyzing volatile or complex fuel mixtures, as it avoids potential sample loss or alteration that might occur during vaporization. Consequently, ATR infrared spectroscopy proves to be an advantageous choice for robust and convenient characterization of liquid fuels, facilitating the accurate and rapid prediction of their properties.

One significant challenge in utilizing ATR-FTIR spectroscopy for fuel property prediction is the need for extensive spectral measurements of fuels and fuel surrogate blends for effective chemometric analysis. This challenge is addressed in this paper by providing a large experimental dataset of ATR-FTIR spectral measurements for RON/MON/DCN prediction and by introducing several data enhancement techniques. These include data imputation using synthetic spectra blending and pseudo-labeling to address missing data, semi-supervised synthetic data generation to expand the dataset size, and consistency enforcement through unsupervised data augmentation to improve model robustness and generalization, especially for complex real fuels. A rigorous out-of-distribution evaluation methodology is employed to assess the model's reliability applicability and generalization beyond the training data.

\section{Methodology}
\subsection{Dataset curation}

A comprehensive dataset of fuel compositions and properties was curated from literature sources to train and evaluate the predictive models developed in this study. The final dataset consists of 757 entries, including 104 pure components, 518 surrogate blends, and 135 real fuels. Each entry is labeled with at least one of the target properties (RON, MON, or DCN) based on experimental measurements in accordance with ASTM D2699, D2700, D6890, D7170, or D7668 standards. For consistency and to mitigate inconsistencies from fuel purity or observer bias, reported values were averaged across sources when multiple values were available for the same sample.

For pure components, property values were compiled from established compendia and literature sources \cite{Lynch, Project45, yanowitz2017compendium, yanowitz2011utilization, naegeli1989measurement, mccormick2017selection, NN-GCM, FunctionalGroup, RD-15-00161, egloff1941octane, szybist2020update, schweidtmann2020graph, guo2017predicting, hunwartzen1982modification, neumann2024fuel}.

Surrogate blends were assembled from studies that formulated and experimentally validated multicomponent fuel mixtures designed to represent gasoline, diesel, or jet fuel properties. These blends were drawn from literature sources \cite{Lynch, SM4, SM3, AGDCN, 21inHanson, 22inHanson, surrogate, waqas2017auto, waqas2016blending, angikath2020investigating, jameel2021predicting, fioroni2022chemical, SM2, kim2019two, SM1, by-mole}, and their ATR-FTIR spectra were measured experimentally in this work.

Real fuels include a broad array of conventional and synthetic fuels, such as methanol-to-gasoline and ethanol-to-gasoline blends, and were selected from sources reporting compositional data and measured RON, MON, or DCN values under ASTM-compliant conditions \cite{SM4, waqas2017auto, waqas2016blending, angikath2020investigating, jameel2021predicting, naser2018ignition, nicolle2019autoignition, CRC, NEWDIESEL2, AGMFG2, lee2017autoignition, aljohani2024impact}. These real fuels often contain oxygenate additives such as alcohols, ethers, and furans. ATR-FTIR spectra for all real fuels were collected in this study.

 All spectral measurements used in model development were obtained using the experimental procedure described in Section~\ref{sec:expdetails}. Property data utilized in this work and the corresponding measured ATR-FTIR spectra are included in supplementary materials.

\subsection{Experimental details}
\label{sec:expdetails}
\subsubsection{Sample preparation}

Mixtures of $1\ mL$ volume were prepared using a ThermoScientific Pipette (100-1000 $\mu$L). All samples were preserved in $5\ mL$ sterile glass vials. All pure components used in this study have a purity $>99\%$. Temperature during sample preparation and spectra collection was maintained at 20 ± 1.0 °C.

\subsubsection{Spectra collection}

The spectra were collected using a ThermoFisher Nicolet iS10 FTIR with a Smart Multi-Bounce HATR (Horizontal Attenuated Total Reflection) accessory. The ATR accessory crystal is a $45^{\circ}$ ZnSe that produces 10 reflections. The ATR accessory also includes a high-pressure clamp accessory, which was used to keep the sample closed during the measurement time to prevent evaporation. The spectra were recorded by averaging 16 scans between $4000-650\ cm^{-1}$ with a data spacing of $0.241\ cm^{-1}$. The spectral resolution is set to $2\ cm^{-1}$ and the interferometer mirror speed is set to $2.8\ mm/s$. 

In this study, we report the absorbance spectra as:

\begin{equation}
    A = - log_{10}\left(\frac{r_\%}{100}\right) = - log_{10}\left(\frac{\frac{I_{abs}}{I_0}}{100}\right) = - log_{10}\left(\frac{1- \frac{I_{detector}}{I_0}}{100}\right),
\end{equation}

where $r_\%$ is the reflectance percentage, $I_0$ is the intensity coming from the source, $I_{abs}$ is the absorbed intensity and $I_{detector}$ is the intensity measured in the detector. For more information on the experimental procedure and its uncertainties please refer to \cite{Non-ideal}.

\subsection{Preprocessing}

Minimal preprocessing was done to the raw measured absorbance spectra. A brief description of data preprocessing steps is given below. For further details please refer to \cite{Non-ideal}.
\subsubsection{Baseline correction}

 The first preprocessing step in this study was to subtract the minimum value of each absorbance spectrum from all spectral points in order to reset the baseline to zero absorbance. This is done to eliminate small negative absorbance values that are caused by scattering, unwanted reflection, temperature change, or instrument abnormalities. Equation \ref{eq:BLcor} shows the baseline corrected signal $\Tilde{A}(\lambda)$:
\begin{equation}\label{eq:BLcor}
\begin{split}
    \Tilde{A}(\lambda) &= A(\lambda) - A_{min}
\end{split}
\end{equation}

\subsubsection{Saturation correction}
The second preprocessing step was to set an upper absorbance limit of 2 to deal with saturation. Saturation occurs when the absorbance at a certain wavelength is too high that most of the incident light gets absorbed leading to an unreliable measurement. The saturation adjustment can be described as follows:

\begin{equation}\label{eq:SATcor}
\begin{split}
    A_{sat} = 2 &\longrightarrow r_{\%sat} = 100\times10^{-2}= 1,\\
    \Tilde{r}_\%(\lambda) &= \max\left(r_\%(\lambda),1\right).
\end{split}
\end{equation}

Thus all the spectra used in this study are capped between [0-2] absorbance values. 

\subsection{Data enhancement}\label{sec:enhancement}
This section details data enhancement techniques used to address limitations in the reported dataset. These techniques are data imputation through synthetic spectra blending and pseudo-labeling, semi-supervised synthetic data generation, and consistency enforcement through unsupervised data augmentation. These offer distinct approaches to improving data quality and quantity for machine learning. While all aim to mitigate challenges related to data scarcity or quality, their methods and objectives differ substantially. Data imputation fills in missing values within the existing dataset, preserving the original data structure but replacing incomplete entries with estimated values. Data augmentation, on the other hand, creates modified versions of existing data points, expanding the dataset and introducing variations while maintaining the underlying distribution. It enhances the model's ability to generalize, but does not directly address missing data. Finally, synthetic data generation creates entirely new data points that mimic the characteristics of the original dataset but are independent of the original data. This is used to address issues of small datasets or even to generate data for specific scenarios not represented in the original data. The following subsections elaborate on each of these techniques and their implementation in this work individually.

\subsubsection{Data imputation: synthetic spectra blending and pseudo-labeling}

Data imputation is a crucial preprocessing technique in machine learning that addresses the common challenge of missing values in datasets. When working with small datasets, discarding incomplete observations limits the ability of machine learning models to generalize and capture complex relationships within the data. To address missing data in this work, two strategies are employed: synthetic spectra blending and pseudo-labeling.

First, synthetic spectra blending is utilized for data points where the fuel's composition and target labels (RON, MON, or DCN) are known but experimental spectral measurements are missing. This involves generating the missing spectra by synthetically blending the spectra of its components. For real fuels, a representative weighted sum of class-averaged spectra ($\overline{A}_{class}$) of 21 paraffins, 5 olefins, 10 naphthenes, 12 aromatics, and 54 oxygenates is used following the procedure described in \cite{OctanePaper}. Oxygenates were excluded from spectral generation of diesels because the oxygen content is less than 0.1 (molar\%) in all samples. The weights, used in the following equation, come from detailed hydrocarbon analyses (DHA) \cite{AandC,FandG,FACEI} according to ASTM D6730 \cite{ASTM6730} and D6733 \cite{ASTM6733} or combined $^1$H and $^{13}$C NMR analysis \cite{NEWDIESEL,NEWDIESEL2}.

\begin{equation}
A^{real}=\sum_{i=1}^{N} w_i \overline{A}_{i}
\end{equation}
where $A^{real}$ represents the synthetically generated absorbance spectrum, $w_i$ represents the weight of the $i^{th}$ class-averaged absorbance spectrum, $\overline{A}_{i}$ is the class-averaged absorbance spectrum of the $i^{th}$ class, and $N$ is the total number of classes considered.

For surrogate blends, spectral generation was performed by blending the spectra of surrogate components by mole fraction. The ideal blended absorbance ($A^{ideal}$) is calculated using the Beer-Lambert relation, assuming linear blending by mole fraction:

\begin{equation}\label{eq:linearbymole}
    A^{ideal} = \sum_{i=1}^N x_{i}A_{i}^{pure}
\end{equation}

where $x_i$ is the molar concentration of the $i^{th}$ component, N is the total number of components, and $A^{pure}_i$ is the experimentally obtained absorbance spectrum of the $i^{th}$ pure component. However, this ideal model does not fully account for intermolecular interactions in the condensed phase. Therefore, we employed the non-linear blending equation from \cite{Non-ideal} to correct for these non-ideal effects. Here, we use the general non-linear blending equation developed in \cite{Non-ideal}:

\begin{equation}\label{eq:A_nonideal}
    A^{non-ideal} = \left(1-\sum_kf_k\right) 
    \underbrace{\left(\sum_i  \left[\frac{x_i}{1-\sum_kx_k}\right]A_i\right)}_\text{Linear}+ \underbrace{\sum_kf_kA_k}_\text{non-Linear}
\end{equation}

where $k$ indexes components exhibiting non-linear behavior and $f_k$ represents the calibration curve for the $k^{th}$ component.

Second, for data points where experimental spectra are available but target labels were missing or partially missing, we employed pseudo-labeling. In this semi-supervised learning approach, a model is first trained on a limited set of labeled data points. This initial model is then used to predict target labels for unlabeled data, generating pseudo-labels. These pseudo-labels, while not definitive ground-truth values, provide valuable information about the underlying data distribution. The model is then retrained on the combined original labeled dataset and pseudo-labeled dataset, improving its ability to generalize. 

\subsubsection{Semi-supervised synthetic data generation}\label{sec:syn_data}
Semi-supervised data generation combines both data imputation strategies described above to generate new data points that have artificial spectra and pseudo-labels. This approach allows for expanding our dataset size. The artificial spectra are created by blending spectra of the pure components in our dataset weighted by randomly generated mole fractions. Labels are then obtained by pseudo-labeling as described above. The aim of this approach is to enhance the diversity and quantity of data available for analysis, ultimately improving the robustness of the developed model.

\subsubsection{Unsupervised data augmentation}
\begin{figure}[!htbp]
\centering
  \includegraphics[width= 0.9\textwidth]{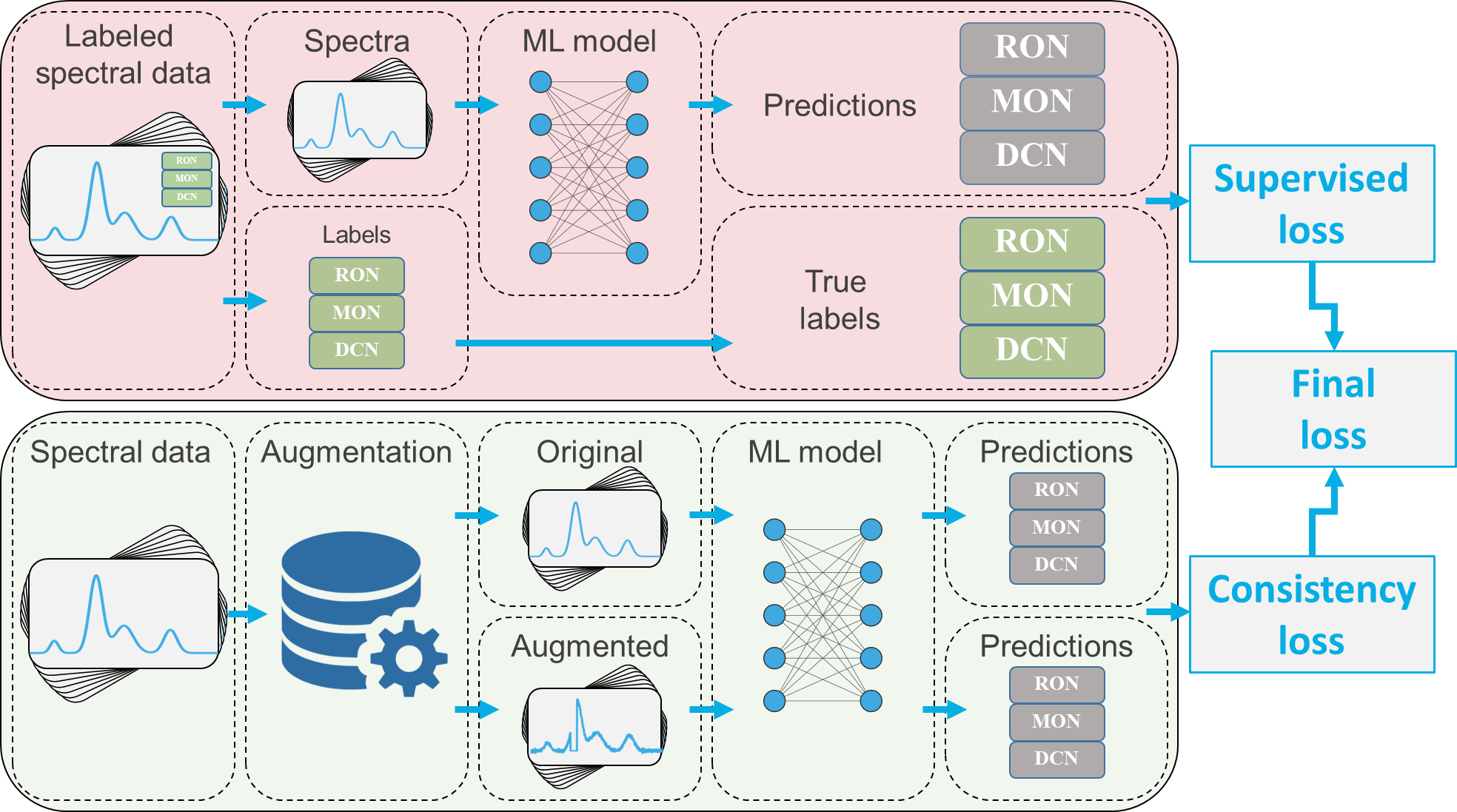}
\caption{Schematic explaining semi-supervision using unsupervised data augmentation.}
\label{fig:fuelprop_cl}
\end{figure}
\begin{figure}[!htbp]
\centering
  \includegraphics[width= 0.7\textwidth]{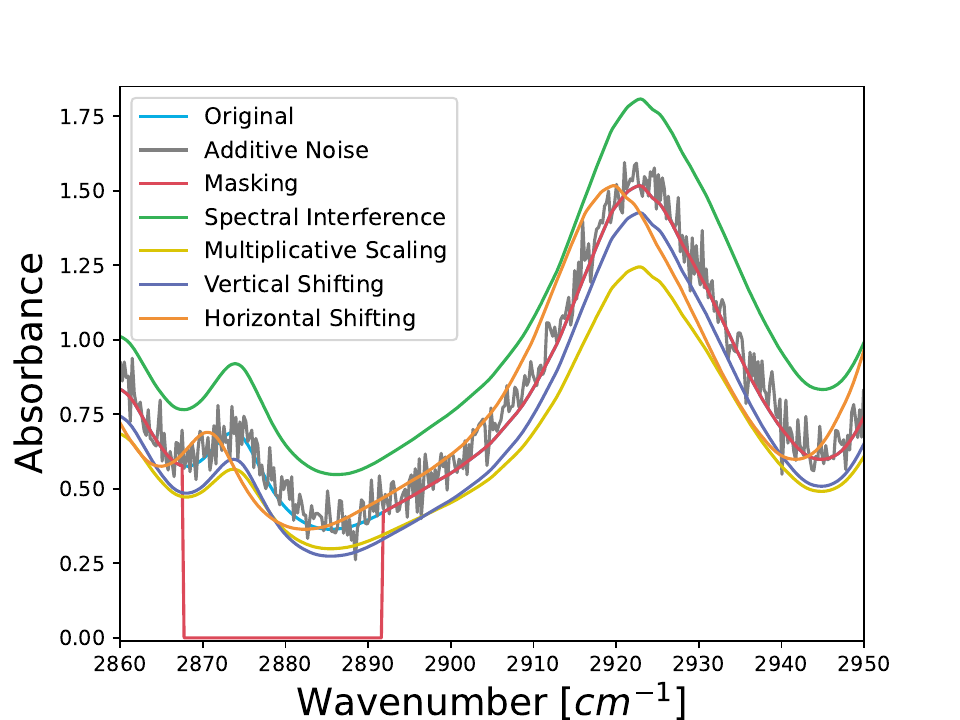}
\caption{Examples of data augmentation techniques applied to spectra.}
\label{fig:augmentations}
\end{figure}

This work addresses the challenge of generalizing a model trained on surrogate fuels to complex real fuels. The diverse and varying compositions of real fuels hinder direct generalization. To improve model robustness and generalization, an unsupervised data augmentation (UDA) framework is adapted here \cite{xie2020unsupervised}. This leverages consistency training, minimizing the discrepancy between predictions made on original and augmented spectral data. Label-preserving augmentations, including masking, horizontal/vertical shifting, additive noise, and simulated spectral interference, are utilized in this work. These augmentations aim to create diverse yet realistic spectral variations that a robust model should generalize to. The specific augmentations and their parameters are detailed below (see Figure \ref{fig:augmentations} for illustrative examples). A schematic of the consistency training process is provided in Figure \ref{fig:fuelprop_cl}.

\textbf{Multiplicative scaling:} This augmentation scales the spectral values by a random factor, simulating variations in sensor sensitivity or sample concentration. The multiplier is drawn from a uniform distribution and applied across all spectral points.

\begin{algorithm}[H]
\caption{Multiplicative Scaling Augmentation}
\label{alg:multip}
\begin{algorithmic}[1]
\Require Spectra tensor $X$, scaling factor range $s$
\For {each spectrum $x_i$ in $X$}
\State Sample multiplier $\alpha \leftarrow 1 + U(-s, s)$
\State Multiply $x_i \leftarrow x_i \times \alpha$
\EndFor
\end{algorithmic}
\end{algorithm}

\textbf{Masking:} Masking augmentation introduces random zero-value segments into the spectra to simulate missing or corrupted regions due to sensor noise or occlusions. A random segment length, determined by a predefined maximum mask size, is selected, and a continuous region within the spectrum is set to zero.

\begin{algorithm}[H]
\caption{Masking Augmentation}
\label{alg:masking}
\begin{algorithmic}[1]
\Require Spectra tensor $X$, mask size $m$
\For {each spectrum $x_i$ in $X$}
\If {random probability $> 0.5$}
\State Choose random patch size $p \leftarrow U(0, m)$
\State Choose random index $j \leftarrow U(0, |x_i| - p)$
\State Set $x_i[j:j+p] \leftarrow 0$
\EndIf
\EndFor
\end{algorithmic}
\end{algorithm}

\textbf{Spectral interference:} This augmentation introduces spectral interference by blending the spectra with a set of reference spectra using randomly weighted contributions. The interference can be either additive or subtractive.

\begin{algorithm}[H]
\caption{Spectral Interference Augmentation}
\label{alg:interference}
\begin{algorithmic}[1]
\Require Spectra tensor $X$, reference spectra $R$, interference scale $c$
\For {each spectrum $x_i$ in $X$}
\State Sample weights $w \leftarrow U(0,1)$
\State Normalize weights $w \leftarrow w / \sum w$
\State Compute interference $I \leftarrow w^T R \times c$
\State Sample sign $s \leftarrow \pm 1$
\State Apply interference $x_i \leftarrow x_i + sI$
\EndFor
\end{algorithmic}
\end{algorithm}

\textbf{Additive noise:} Additive noise augmentation introduces random noise sampled from Gaussian and uniform distributions to simulate sensor noise. The noise is scaled by a predefined noise level.

\begin{algorithm}[H]
\caption{Additive Noise Augmentation}
\label{alg:noise}
\begin{algorithmic}[1]
\Require Spectra tensor $X$, noise level $n$
\For {each spectrum $x_i$ in $X$}
\State Sample Gaussian noise $g \leftarrow \mathcal{N}(0, n)$
\State Sample Uniform noise $u \leftarrow U(0, n)$
\State Add noise $x_i \leftarrow x_i + g + u$
\EndFor
\end{algorithmic}
\end{algorithm}

\textbf{Horizontal shifting:} Horizontal shifting simulates misalignment or spectral calibration errors by shifting the spectral values left or right. The vacant positions created by the shift are filled with zeros.

\begin{algorithm}[H]
\caption{Horizontal Shift Augmentation}
\label{alg:h-shift}
\begin{algorithmic}[1]
\Require Spectra tensor $X$, shift range $h$
\For {each spectrum $x_i$ in $X$}
\State Sample shift value $s \leftarrow U(-h, h)$
\State Shift spectrum $x_i \leftarrow \text{roll}(x_i, s)$
\If {$s > 0$}
\State Set $x_i[:s] \leftarrow 0$
\ElsIf {$s < 0$}
\State Set $x_i[s:] \leftarrow 0$
\EndIf
\EndFor
\end{algorithmic}
\end{algorithm}

\textbf{Vertical shifting:} This augmentation introduces a small constant shift to the spectral values, mimicking variations in baseline intensities. A random value within a predefined range is added uniformly across the spectrum.

\begin{algorithm}[H]
\caption{Vertical Shift Augmentation}
\label{alg:v-shift}
\begin{algorithmic}[1]
\Require Spectra tensor $X$, shift range $v$
\For {each spectrum $x_i$ in $X$}
\State Sample shift value $\beta \leftarrow U(-v, v)$
\State Add shift $x_i \leftarrow x_i + \beta$
\EndFor
\end{algorithmic}
\end{algorithm}

\subsection{Modeling}
\subsubsection{Model architecture}
Convolutional neural networks (CNNs) \cite{lecun2015deep} are a popular choice to extract useful information from spectral data with overlapping features \cite{Lynch}. Since CNNs learn task-specific feature maps, they are expected to yield a more generally applicable approximate function than other architectures.
Our implementation of CNNs comprised three convolutional layers with a kernel size of 32 and stride of 8, followed by fully connected layers that separate into three heads, one for each target property (RON, MON, and DCN). PReLU activation functions \cite{PReLU} were used for the fully connected layers. 
To enhance training stability and generalization, batch normalization was applied after each convolutional layer. Regularization was further enforced using weight decay (L2 penalty) to prevent overfitting. Additionally, an early stopping strategy was employed based on validation loss to mitigate overfitting during training. Additionally, a learning rate scheduler was employed to systematically reduce the learning rate facilitating better convergence during training.
The optimization process utilized the Adam optimizer \cite{Adam} and minimized the mean squared error.

\subsubsection{Model evaluation}\label{sec:model_eval}
The choice of testing methodology for fuel property prediction models is crucial for assessing their reliability and applicability. While many models in the literature are evaluated using a random split of the entire dataset (typically comprising pure components, blends, and real fuels) into training and testing sets, a more rigorous test involves out-of-distribution evaluation, where real fuels are exclusively present in the test set. In this work, real fuels are reserved for testing, while the models were trained on $90\%$ of the pure components and fuel surrogate blends, with the remaining $10\%$ used for validation during training. After training, the models are tested in out-of-distribution mode on real fuels (complex mixtures of hundreds of molecules)

\subsubsection{Hyperparameter optimization}
Hyperparameters of the developed models are tuned using the Optuna optimization framework \cite{akiba2019optuna}. Optuna, employing a Bayesian optimization approach, efficiently explores the hyperparameter space by dynamically reducing the search space based on past evaluations. This leads to a more efficient hyperparameter optimization compared to grid search.

\section{Results and discussion}
\subsection{Dataset details}

The dataset used in this work consists of 757 fuel samples spanning a broad compositional and property range, as summarized in Table~\ref{tab:Fuelpropdata}. It includes 104 pure components, 518 surrogate blends, and 135 real fuels, with measured values for RON, MON, and DCN. Table~\ref{tab:Fuelpropdata} summarizes the distribution of RON, MON, and DCN targets across the 757 samples.

\begin{table}[!htbp]
\centering
\caption{Number of data points per target}
\label{tab:Fuelpropdata}
\begin{tabular}{|c|c|c|c|c|}
\hline
\textbf{} & \textbf{RON} & \textbf{MON} & \textbf{DCN} & \textbf{Total} \\ \hline
\textbf{Pure} & 56 & 61 & 59 & 104 \\ \hline
\textbf{Blends} & 255 & 173 & 306 & 518 \\ \hline
\textbf{Real fuels} & 33 & 32 & 114 & 135 \\ \hline
\end{tabular}%
\end{table}

\subsubsection{Real fuels}

\begin{figure}[!htbp]
\centering
  \includegraphics[width= 0.8\textwidth]{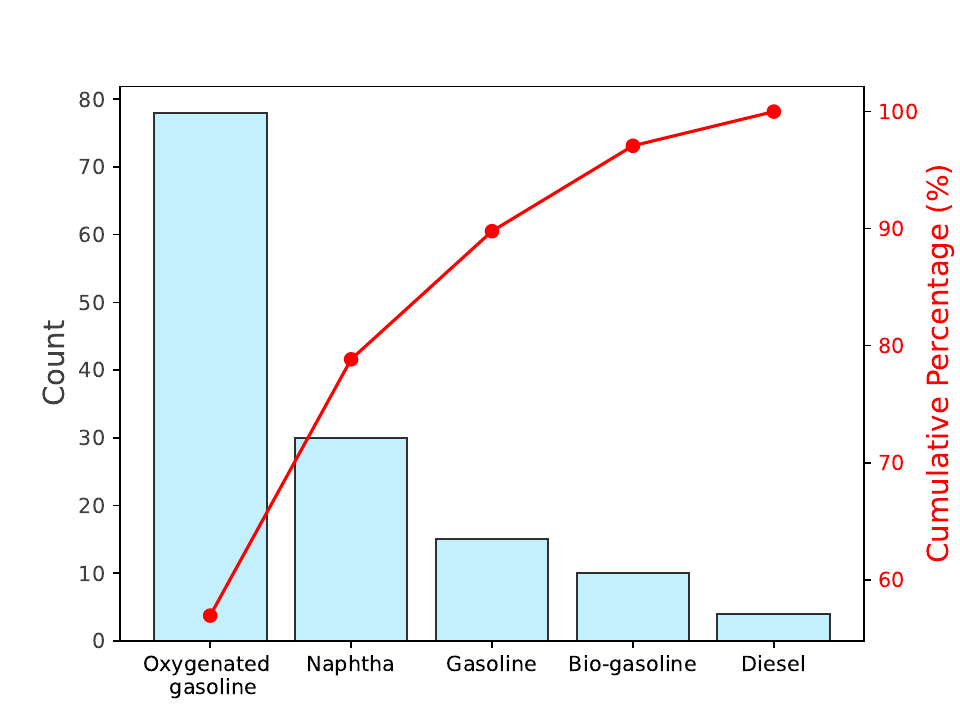}
\caption{Frequency distribution of real fuel types in the dataset.}
\label{fig:reals_hist}
\end{figure}

Real fuels in the dataset are considerably diverse, including gasoline, diesel, and naphtha fuels. These fuels go beyond conventional fossil fuels to include synthetic fuels like methanol-to-gasoline and ethanol-to-gasoline. It is also informed with a variety of oxygenate additives including alcohols, ethers, and furans. This makes it a valuable resource for bench-marking new methods and training strategies. A breakdown of fuel types of real fuels in the dataset is presented in Figure \ref{fig:reals_hist}.

While the distribution of fuel types exhibits an imbalance, the exclusive use of real fuels in the test set (as discussed in Section \ref{sec:model_eval}) mitigates any risk of biasing the model and ensures that this imbalance does not influence model training.

\subsubsection{Pure components and surrogate blends}
To develop a model that is able to generalize to unseen real fuels, surrogate components in our dataset span different classes (n-Paraffin, Iso-paraffin, Olefin, Naphthene, Aromatic, and Oxygenate (PIONA-O); see Table \ref{SurrogateTable}).

\begin{table}[]
\centering
\caption{List of surrogate blend components from representative classes.}
\label{SurrogateTable}
\resizebox{\textwidth}{!}{%
\begin{tabular}{|c|c|c|c|}
\hline
\textbf{Names}                & \textbf{Class}   & \textbf{Names}          & \textbf{Class}   \\ \hline
\textbf{n-heptane}            & n-paraffin       & \textbf{n-decane}       & n-paraffin       \\ \hline
\textbf{n-dodecane}           & n-paraffin       & \textbf{n-hexadecane}   & n-paraffin       \\ \hline
\textbf{n-tetradecane}        & n-paraffin       & \textbf{undecane}       & n-paraffin       \\ \hline
\textbf{iso-cetane}           & isoparaffin       & \textbf{iso-octane}     & isoparaffin      \\ \hline
\textbf{toluene}              & aromatic         & \textbf{n-propylbenzene}& aromatic         \\ \hline
\textbf{1,2,4-trimethylbenzene} & aromatic       & \textbf{tetralin}      & naphthenes       \\ \hline
\textbf{methyl-cyclohexane}   & naphthenes       & \textbf{cyclopentane}   & naphthenes       \\ \hline
\textbf{cyclohexane}          & naphthenes       & \textbf{1-hexene}       & olefin           \\ \hline
\textbf{ethanol}              & oxygenate        & \textbf{methanol}       & oxygenate        \\ \hline
\textbf{propanol}             & oxygenate        & \textbf{butanol}        & oxygenate        \\ \hline
\textbf{2-methylbutan-1-ol}   & oxygenate        & \textbf{2-methylfuran}  & oxygenate        \\ \hline
\textbf{2,5-dimethylfuran}    & oxygenate        & \textbf{prenol}         & oxygenate        \\ \hline
\textbf{diethylether}         & oxygenate        &                         &                  \\ \hline
\end{tabular}%
}
\end{table}

\begin{figure}[!htbp]
\centering
  \includegraphics[width= \textwidth]{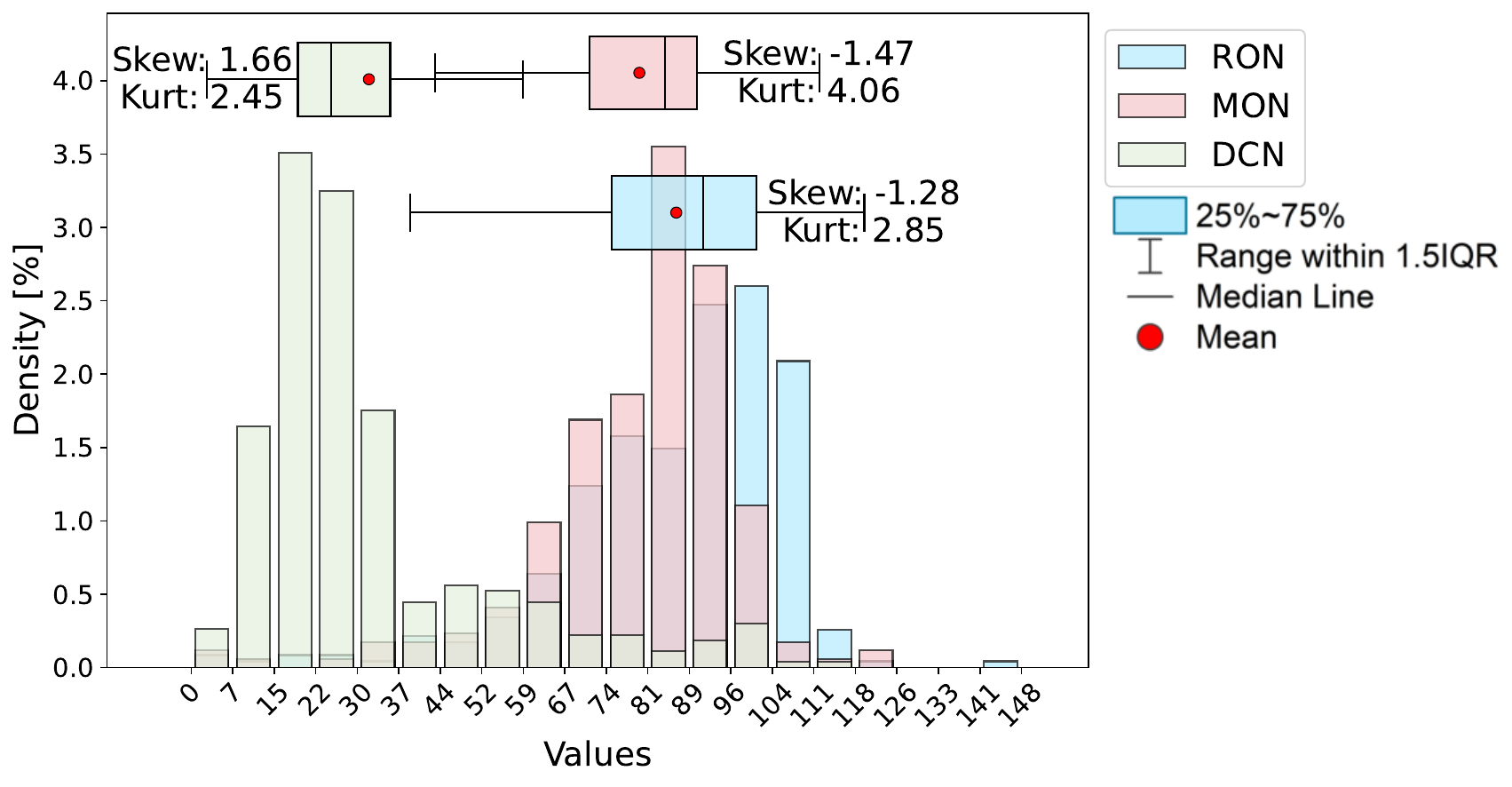}
\caption{Frequency distribution of target values in the training dataset.}
\label{fig:training_dist}
\end{figure}

The training dataset's output values (RON, MON, and DCN) display distinct distributional characteristics, as reflected in their kurtosis and skewness values. RON values have a kurtosis of 2.85 and a skewness of -1.28, indicating a distribution with fewer extreme values compared to a normal distribution and a moderate left skew. MON values show a higher kurtosis of 4.06 and a skewness of -1.47, suggesting the presence of more extreme values and a pronounced leftward skew. DCN values, with a kurtosis of 2.45 and a skewness of 1.66, have fewer extreme values and a strong rightward skew. These characteristics are visualized in Figure \ref{fig:training_dist}. While the presence of outliers may complicate modeling, removing them is not a viable option, as it could potentially eliminate critical data points and reduce the representation of the chemical space being studied. This necessitates model designs robust to such features in the data.

\subsection{ATR-FTIR spectra}

\begin{figure}[!htbp]
  \includegraphics[width= \textwidth]{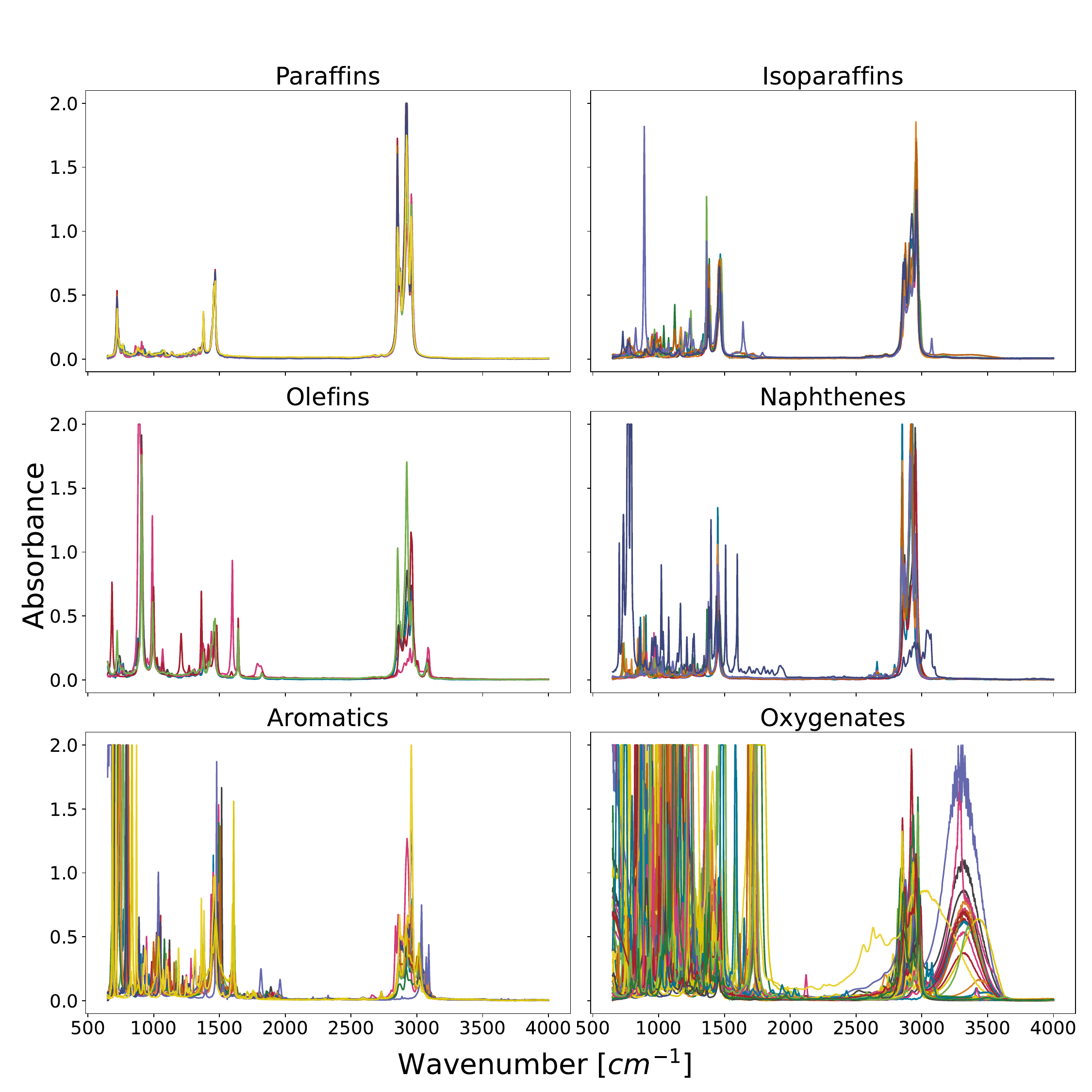}
  \caption{ATR-FTIR spectra of the pure components, categorized by chemical class. Distinct spectral features are observed for each class, highlighting the spectral fingerprints of different chemical structures.}
\label{fig:pure_components}
\end{figure}

Figure \ref{fig:pure_components} displays the spectra of all 104 pure components in our dataset, categorized by their chemical class (n-paraffin, iso paraffin, olefin, naphthene, aromatic, and oxygenate). Key spectral features characteristic of each class are apparent, providing a valuable reference for interpreting the spectra of more complex fuel mixtures.

\begin{figure}
\centering
  \includegraphics[width= 0.7\textwidth]{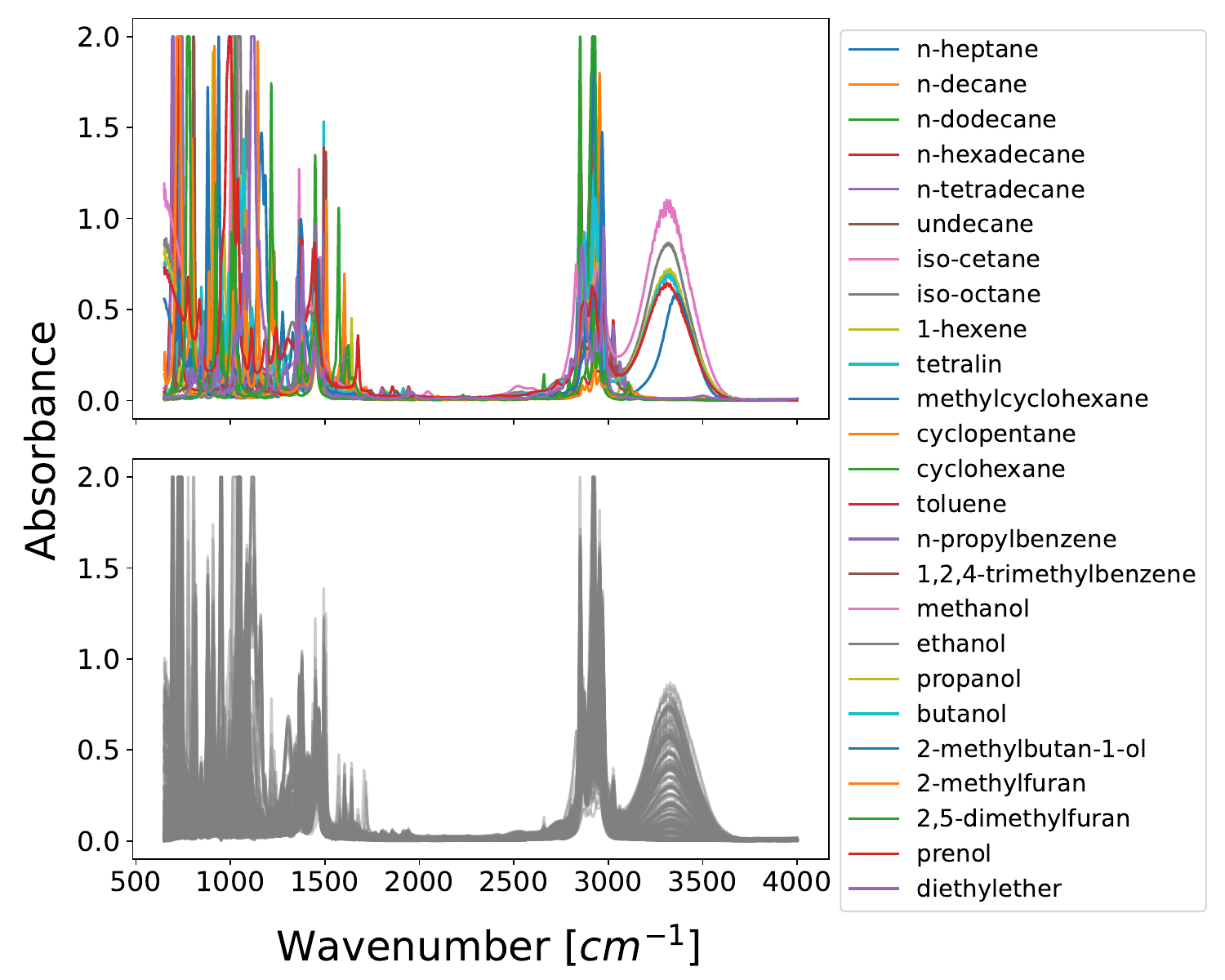}
  \caption{ATR-FTIR spectra of fuel surrogate blends.  The top panel shows spectra of individual pure components, while the bottom panel displays spectra of representative blends (gray lines).}
\label{fig:blends}
\end{figure}

The surrogate fuels from \cite{Non-ideal} were supplemented with additional 287 spectra (all 518 blends are shown in Figure \ref{fig:blends}). 

\begin{figure}
\centering
  \includegraphics[width= 0.8\textwidth]{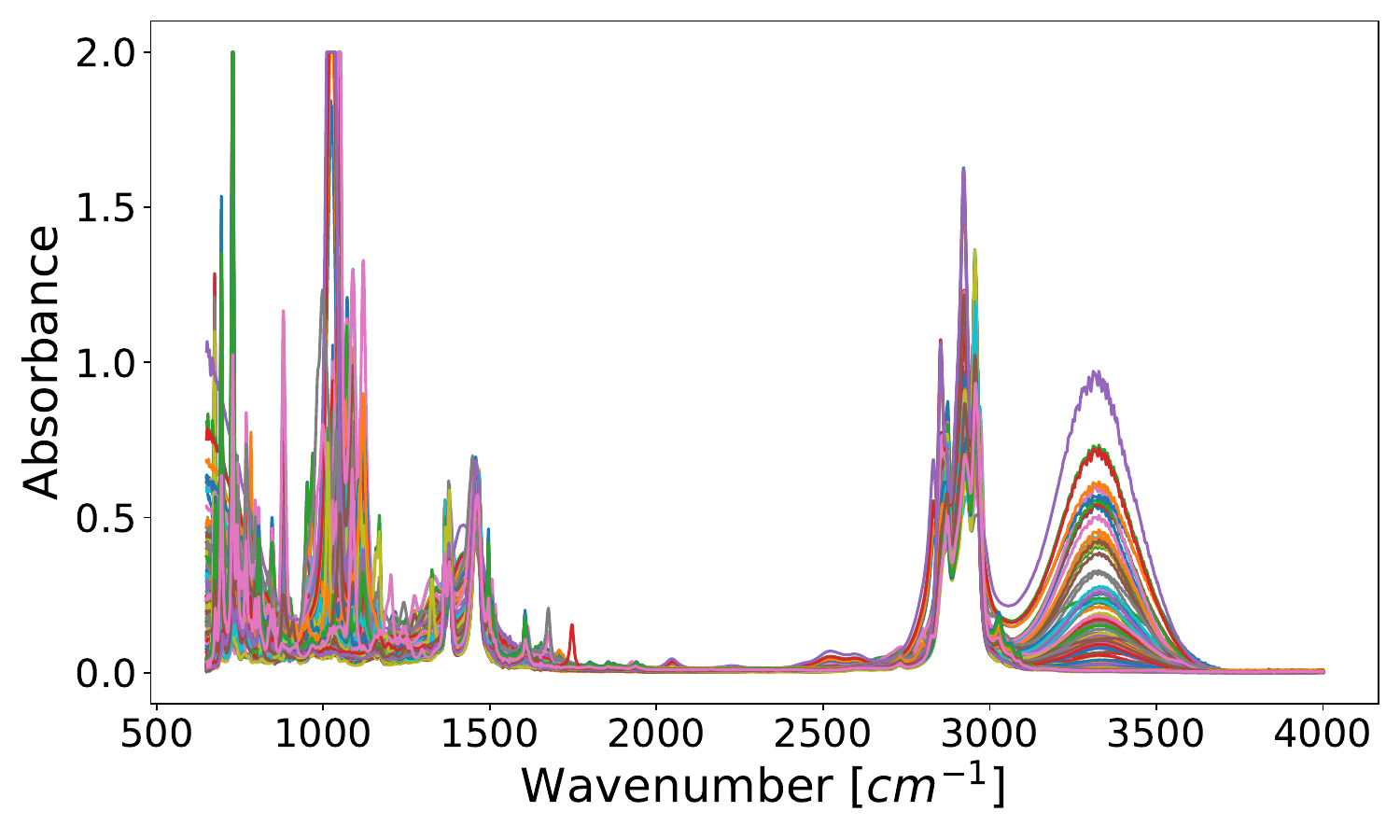}
  \caption{ATR-FTIR spectra of 20 real fuels and their 117 blends.}
\label{fig:reals}
\end{figure}

Figure \ref{fig:reals} presents the spectra of 20 real fuels and 115 real fuel blends. Comparison of Figures \ref{fig:pure_components}, \ref{fig:blends}, and \ref{fig:reals} reveals a strong degree of spectral similarity between the training data (pure components and surrogate blends) and the test data (real fuels and their blends). This spectral consistency suggests that the surrogate blends effectively capture the key spectral characteristics of real fuels, providing confidence in the robustness and generalizability of any models trained on this dataset.

\begin{figure}[!htbp]
\centering
\begin{minipage}{0.48\textwidth}
  \includegraphics[width=\textwidth]{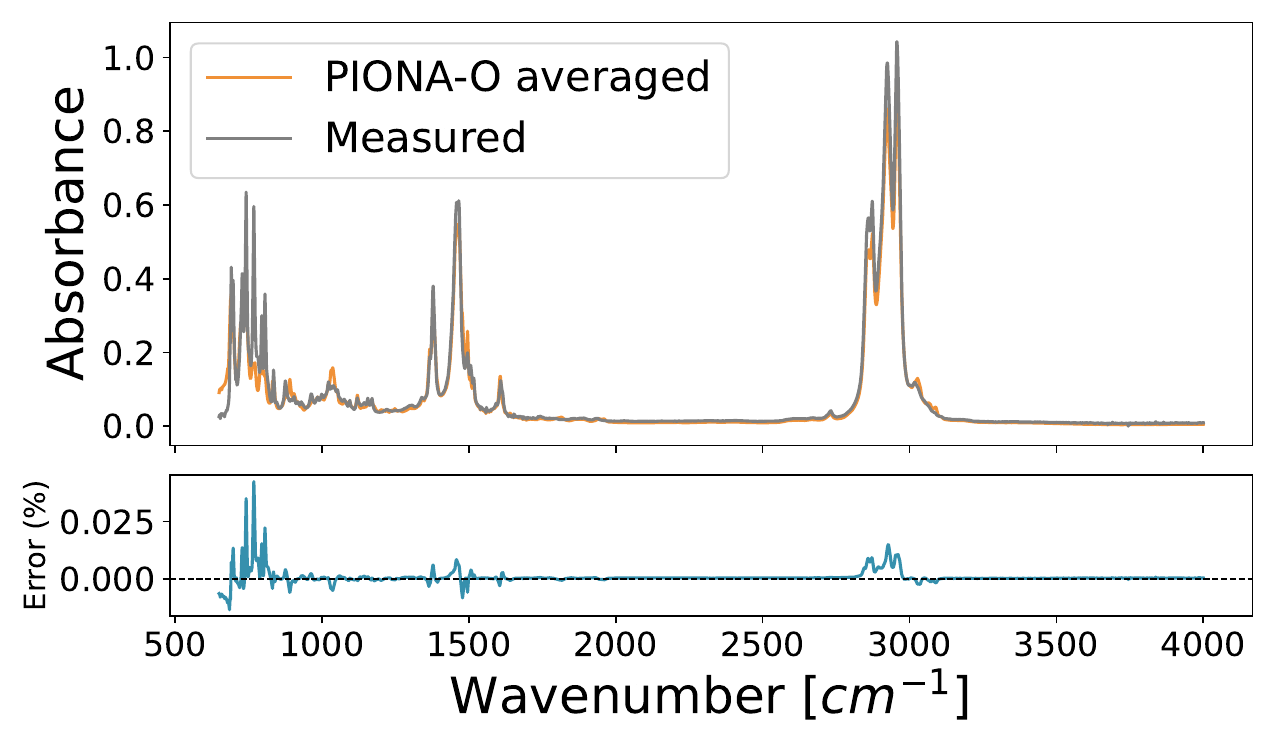}
  \caption{Comparison of measured and synthetically generated mid-infrared spectra for FACE J gasoline. The weighted error, calculated as in Eqn. \ref{eq:weighted_error}, is shown below. The synthetic spectrum is generated by blending PIONA-O class-averaged spectra.
  \newline
  }
  \label{fig:FACEJ}
\end{minipage}
\hfill 
\begin{minipage}{0.48\textwidth}
  \includegraphics[width=\textwidth]{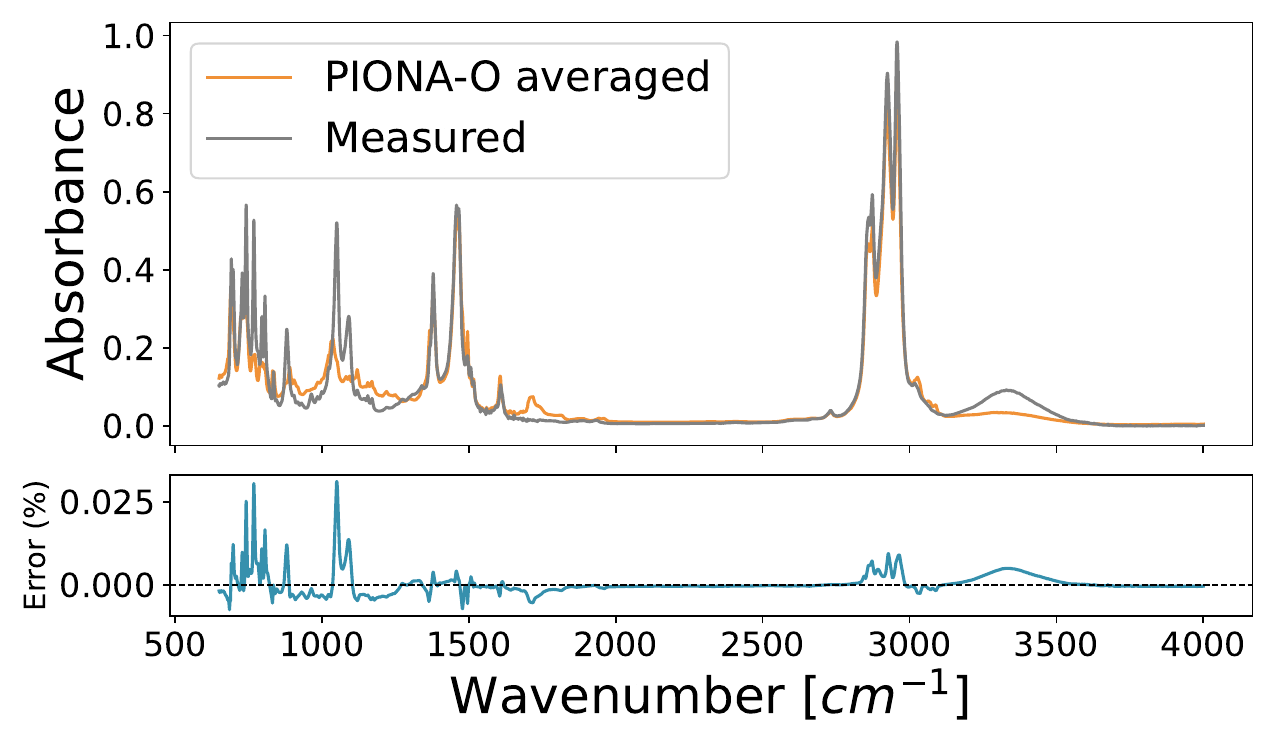}
  \caption{Comparison of measured and synthetically generated mid-infrared spectra for FACE J gasoline blended with 10\% methanol by volume. The weighted error, calculated as in Eqn. \ref{eq:weighted_error}, is shown below. The synthetic spectrum is generated by blending PIONA-O class-averaged spectra.}
  \label{fig:FACEJ10}
\end{minipage}
\end{figure}

The ability to generate synthetic spectra using PIONA class-averaged spectra is demonstrated in Figures \ref{fig:FACEJ} and \ref{fig:FACEJ10}. These figures compare measured spectra of FACE J gasoline (Figure \ref{fig:FACEJ}) and a blend of FACE J gasoline with 10\% methanol (Figure \ref{fig:FACEJ10}) to their synthetic counterparts generated by blending the PIONA-O class-averaged spectra. The weighted error, calculated as: \newline
\begin{equation}\label{eq:weighted_error}
Weighted\ Error\% = \frac{(measured - synthetic)}{\sum{measured}} \times 100
\end{equation}
is shown below each comparison. The relatively low weighted errors in both cases indicate that this approach yields synthetic spectra that accurately capture the essential spectral characteristics of real fuels, even with relatively simple blends. This highlights the potential for this method to aid in the development of predictive models for fuel properties.

\subsection{Predictive modeling}

This study evaluates five predictive models for fuel properties using a multi-target convolutional neural network (CNN) architecture as a baseline. Four data enhancement strategies, synthetic spectra blending, pseudo-labeling, synthetic data generation, and consistency enforcement via unsupervised data augmentation (detailed in Section \ref{sec:enhancement}), are individually assessed resulting in four separate models. In addition, a composite model that integrates all four strategies is evaluated. The composite model is referred to here as 'Fuelprop'. All models are evaluated using the out-of-distribution testing methodology described previously, where real fuels are excluded from the training and validation sets and reserved solely for testing to assess generalization performance.

To quantify performance, each model was trained and evaluated 20 times, using different random seeds for initialization. The performance of each model was measured using mean absolute error (MAE) on the test set. The results are presented as boxplots, where the distribution of MAE values across all runs is visualized. In these plots, the box represents the interquartile range (IQR) between the 25th and 75th percentiles, while the whiskers extend to 1.5 times the IQR. The red marker denotes the mean MAE, while the horizontal line inside the box represents the median. By comparing the MAE distributions between the baseline and each strategy, their effectiveness can be assessed in terms of accuracy (lower MAE values) and consistency (reduced variance across runs).

\subsubsection{Data imputation: synthetic spectra blending and pseudo-labeling}
\begin{figure}[!htbp]
\centering
\begin{minipage}{0.48\textwidth}
  \includegraphics[width=\textwidth]{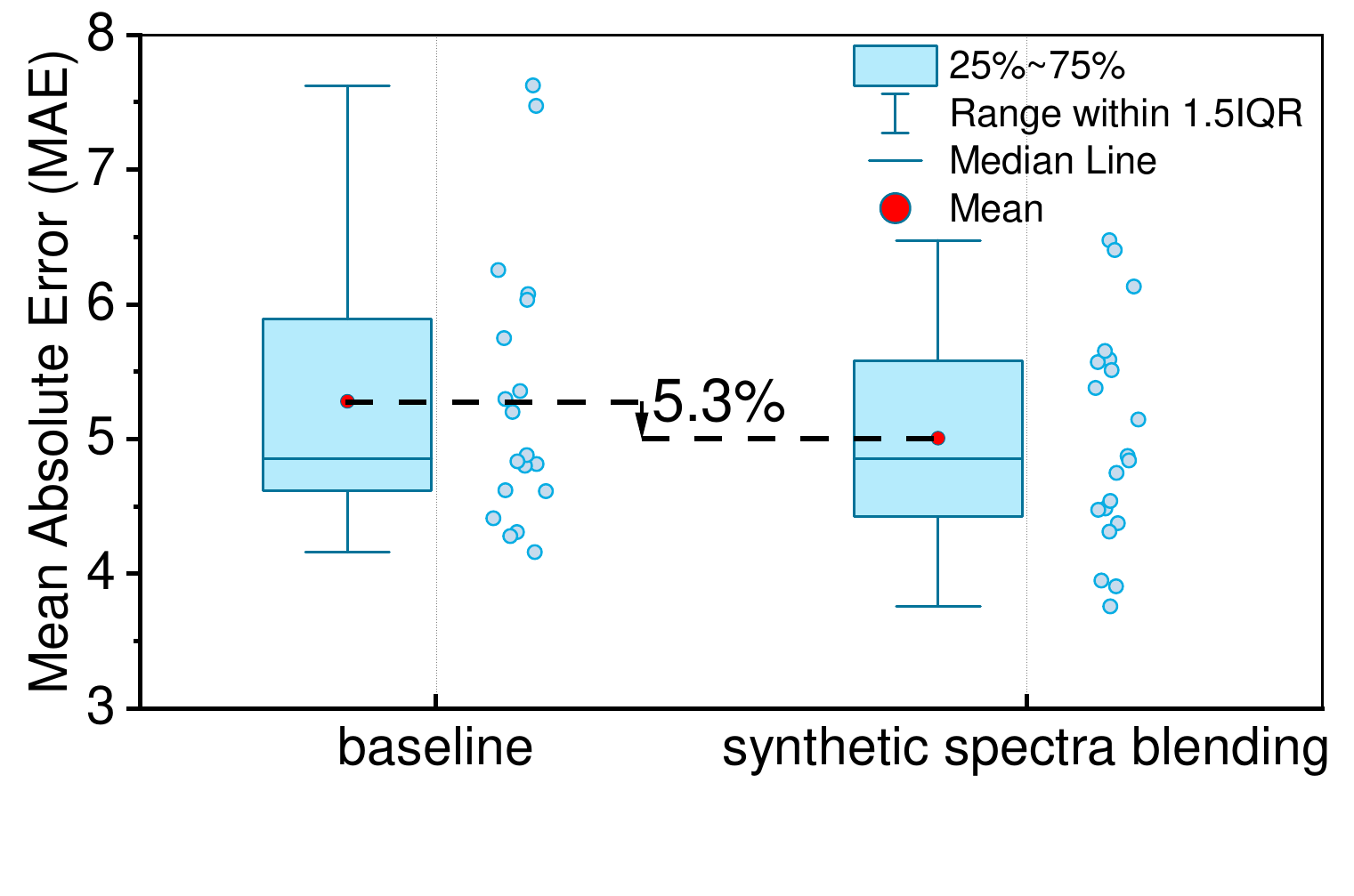}
  \caption{Out-of-distribution performance of a model incorporating synthetic spectra blending compared to the baseline across 20 runs, each with different random seeds.}
  \label{fig:synthetic-spectra}
\end{minipage}
\hfill 
\begin{minipage}{0.48\textwidth}
  \includegraphics[width=\textwidth]{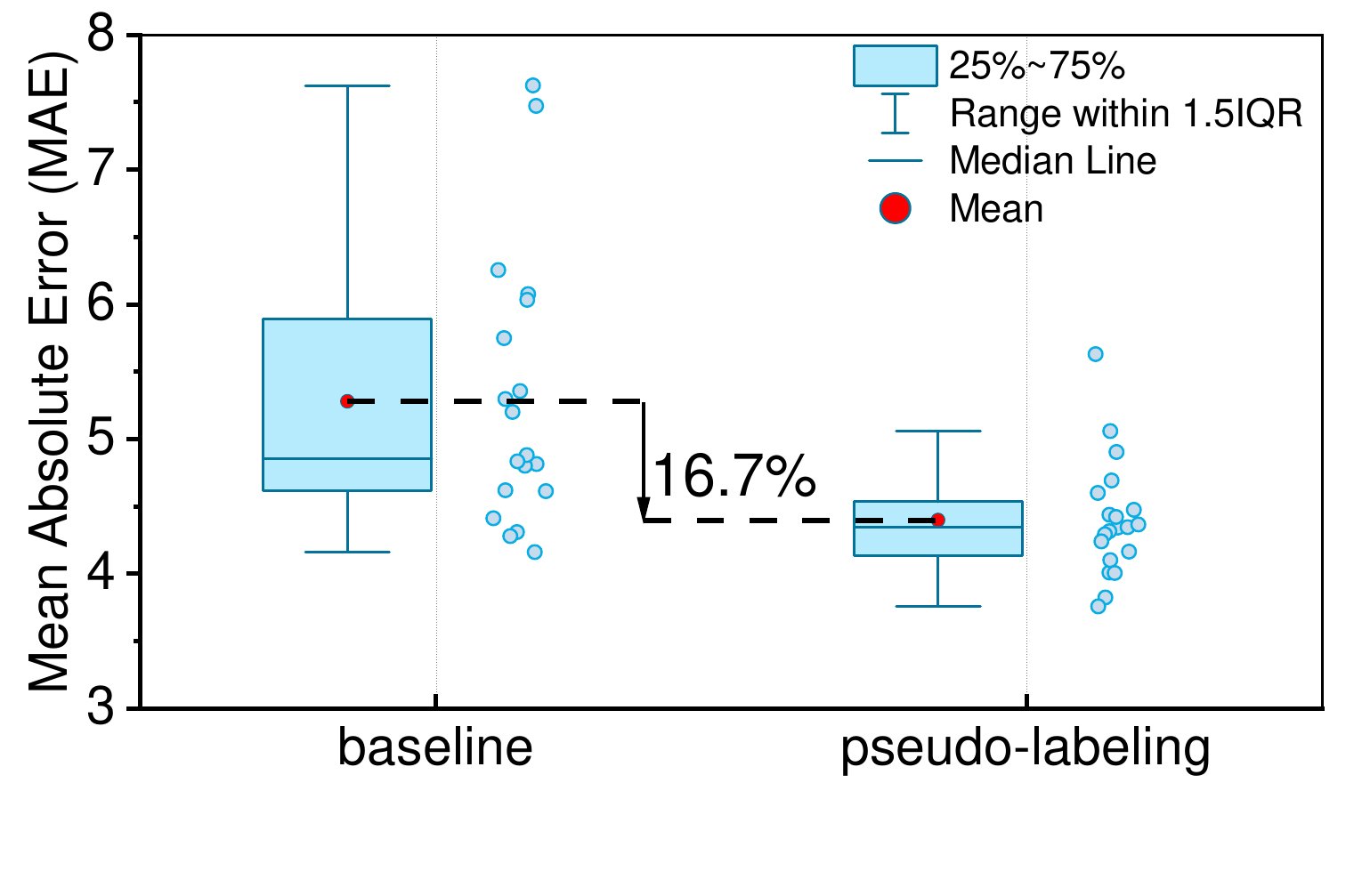}
  \caption{Out-of-distribution performance of a model incorporating pseudo labeling compared to the baseline across 20 runs, each with different random seeds.}
  \label{fig:PL}
\end{minipage}
\end{figure}

The first set of experiments assessed the impact of data imputation, specifically synthetic spectra blending and pseudo-labeling, on predictive accuracy. The results, shown in Figures \ref{fig:synthetic-spectra} and \ref{fig:PL}, illustrate the effect of these imputation techniques compared to the baseline model.

The application of synthetic spectra blending resulted in a 5.3\% reduction in MAE compared to the baseline. By synthetically generating missing spectral data, this approach improved the model’s ability to generalize to real fuel compositions. As shown in Figure \ref{fig:synthetic-spectra}, both the median and mean MAE values decreased, with a slight reduction in variance, indicating improved model stability.

Pseudo-labeling demonstrated a more substantial improvement, leading to a 16.7\% reduction in MAE. By incorporating pseudo-labeled data, the model effectively expanded its labeled dataset. As shown in Figure \ref{fig:PL}, pseudo-labeling not only decreased the mean and median MAE but also significantly reduced the variance, highlighting its role in producing more stable and reliable predictions across different random initializations.

\subsubsection{Semi-supervised synthetic data generation}

\begin{figure}[!htbp]
\centering
\begin{minipage}{0.48\textwidth}
  \includegraphics[width=\textwidth]{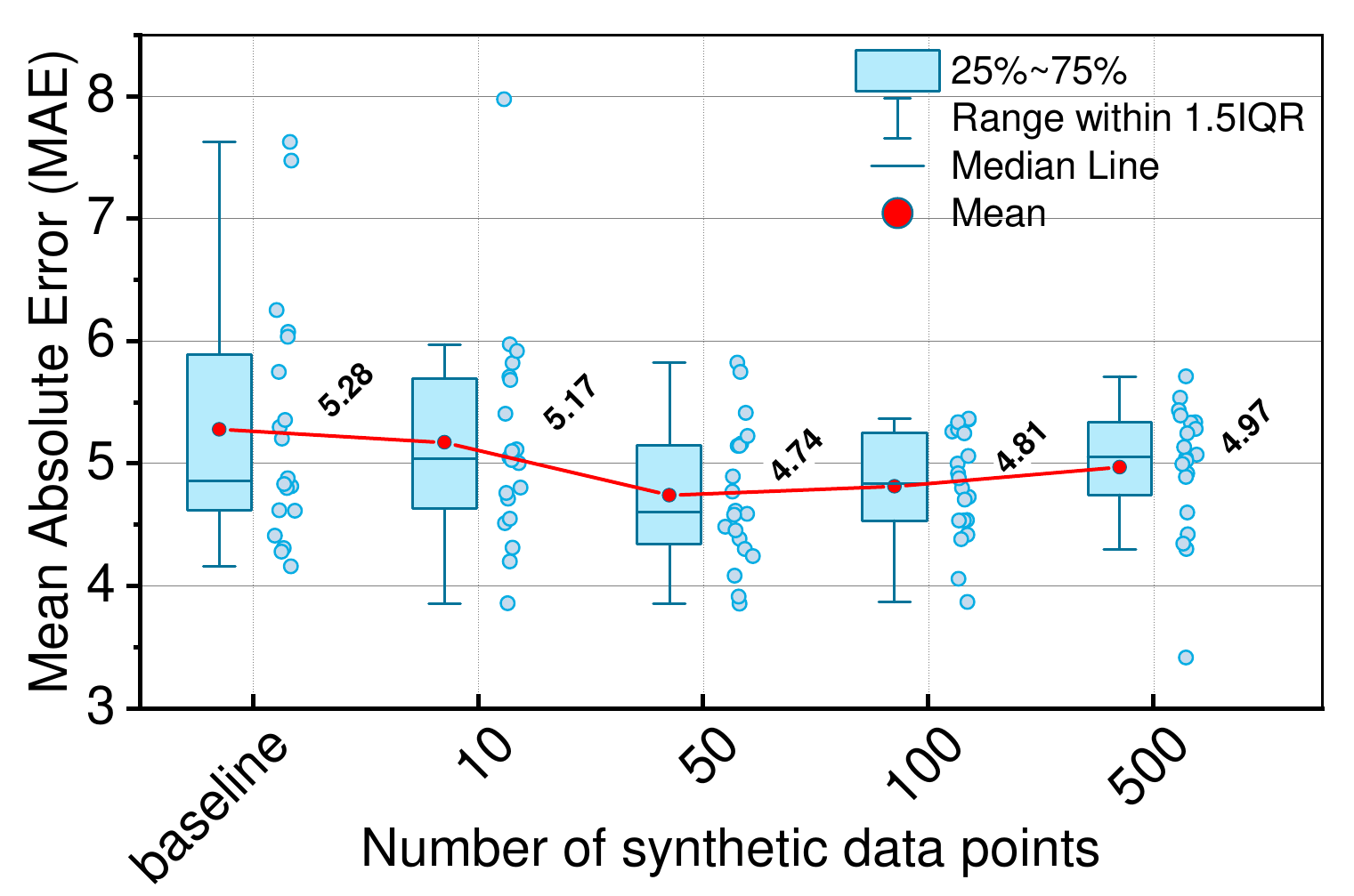}
  \caption{Effect of increasing the number of synthetic data points on model performance. The boxplot compares the out-of-distribution MAE across 20 runs, each with different random seeds.}
  \label{fig:syn-data}
\end{minipage}
\hfill 
\begin{minipage}{0.48\textwidth}
  \includegraphics[width=\textwidth]{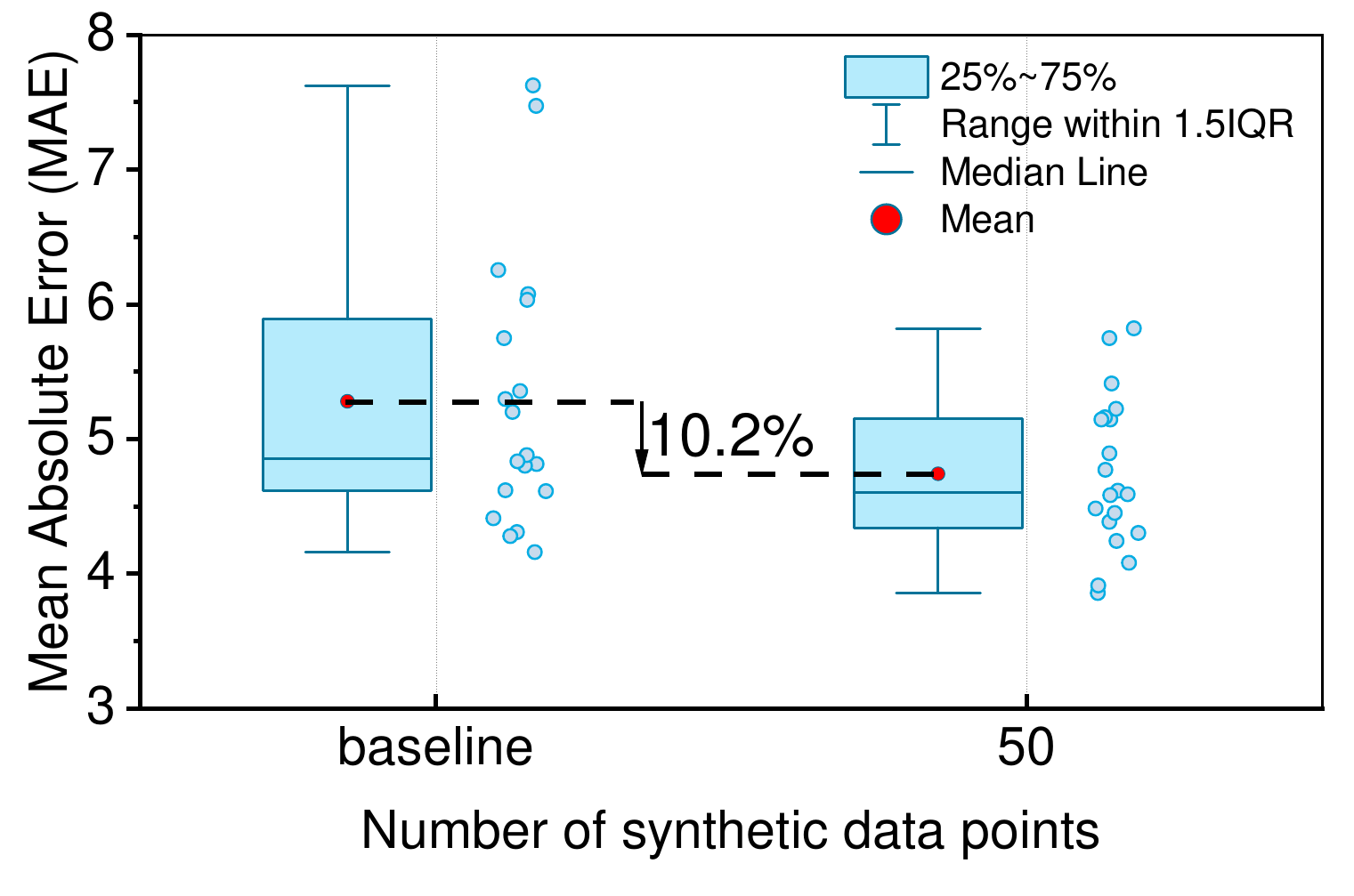}
  \caption{Out-of-distribution performance of the best-performing setting of number of synthetic data points compared to the baseline across 20 runs, each with different random seeds.
  }
  \label{fig:syn-data-2}
\end{minipage}
\end{figure}

The next set of experiments evaluated the impact of semi-supervised synthetic data generation on predictive accuracy. The goal was to assess whether increasing the number of synthetic data points improved the model’s ability to generalize to real fuels. The results, shown in Figures \ref{fig:syn-data} and \ref{fig:syn-data-2}, illustrate how model performance varied as the number of synthetic data points increased and the optimal setting that yielded the highest improvement.

As observed in Figure \ref{fig:syn-data-2}, model performance improves as the number of synthetic data points increases, with a steady reduction in MAE compared to the baseline. However, beyond a certain threshold, the benefits plateau, and further data generation results in diminishing returns. This trend suggests that while synthetic data generation is effective in enhancing predictive accuracy, excessive data generation may introduce noise or redundant information, limiting further gains.

Figure \ref{fig:syn-data} focuses on the best-performing setting for the number of synthetic data points. Under this optimized level, the model achieves a 10.2\% reduction in MAE compared to the baseline.

These findings confirm that semi-supervised synthetic data generation can significantly improve predictive performance, particularly when data is limited.

\subsubsection{Unsupervised data augmentation}
The final set of experiments investigated the impact of consistency enforcement through unsupervised data augmentation on predictive accuracy. Each augmentation type used in this work was examined separately to assess its individual contribution to model generalization.

\begin{figure}[!htbp]
\centering
\begin{minipage}{0.48\textwidth}
  \includegraphics[width=\textwidth]{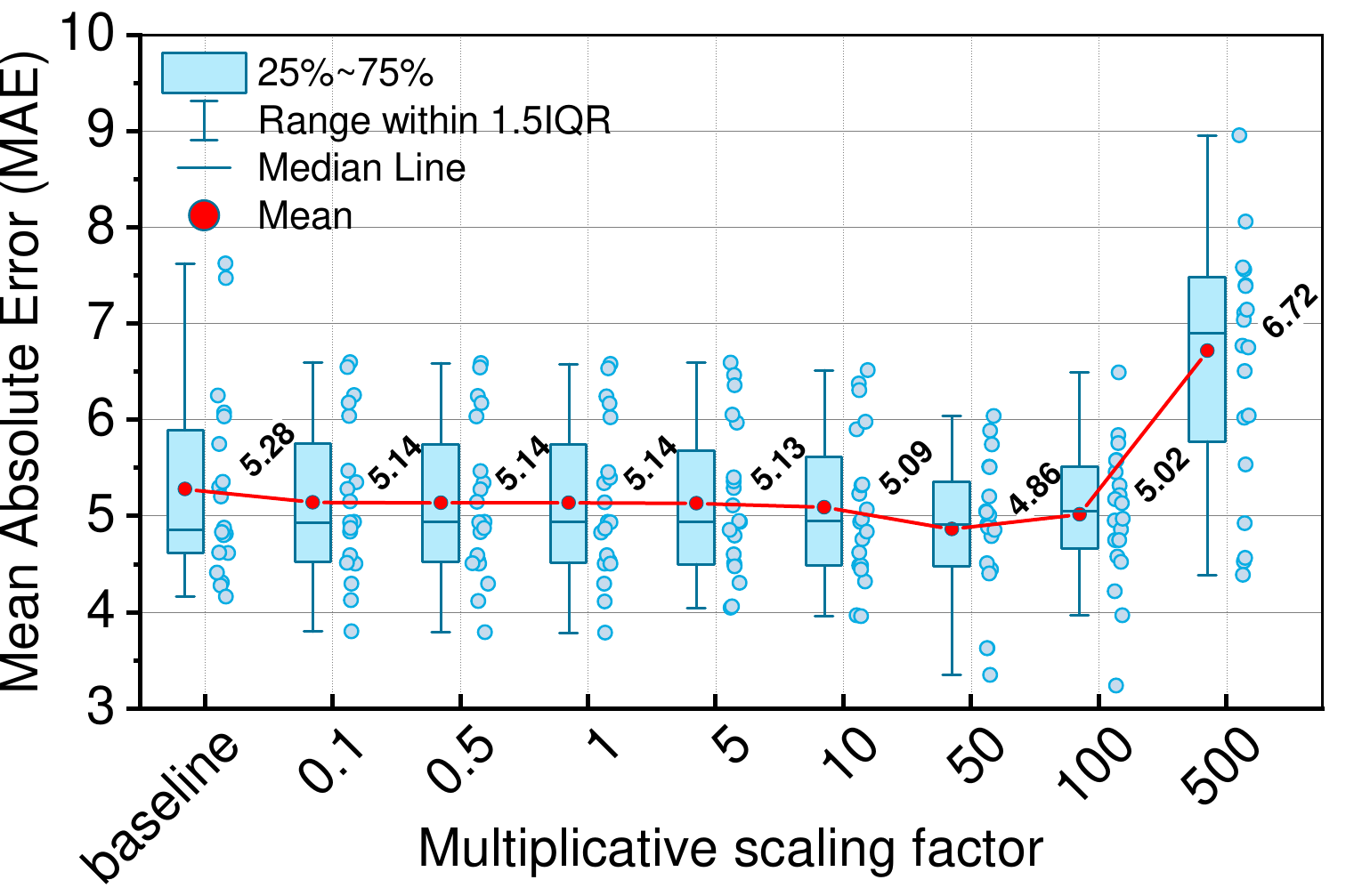}
  \caption{Effect of varying the multiplicative scaling factor range for multiplicative scaling augmentation on out-of-distribution performance. The boxplot compares the out-of-distribution MAE across 20 runs, each with different random seeds.}
  \label{fig:multiplicative-scaling-multiplier}
\end{minipage}
\hfill
\begin{minipage}{0.48\textwidth}
  \includegraphics[width=\textwidth]{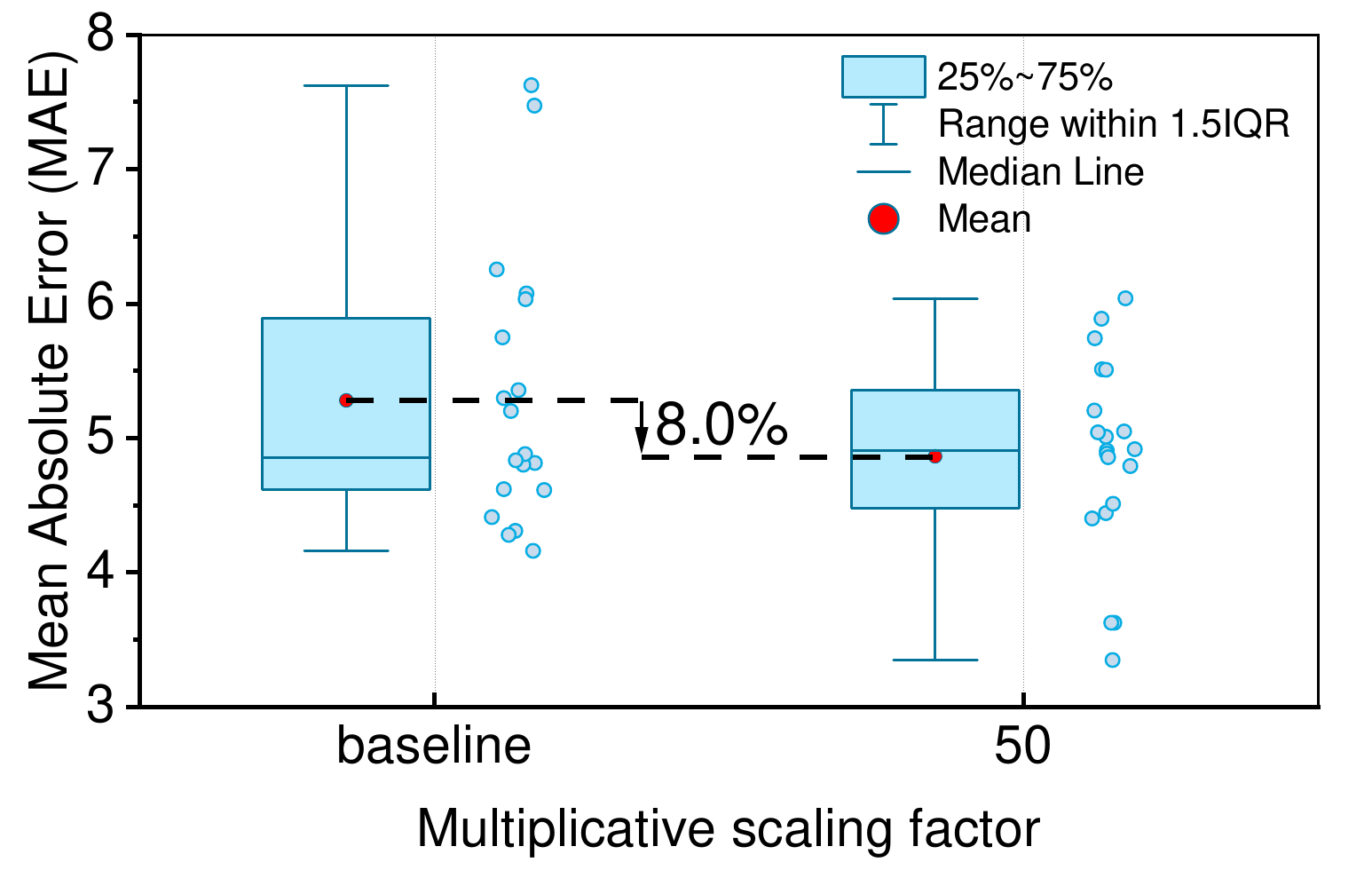}
  \caption{Out-of-distribution performance of the best-performing multiplicative scaling augmentation setting (factor range = ±50) compared to the baseline across 20 runs, each with different random seeds.\newline}
  \label{fig:multiplicative-scaling-best}
\end{minipage}
\end{figure}

\textbf{Multiplicative scaling:} Figure \ref{fig:multiplicative-scaling-multiplier} shows the effect of varying the range of the multiplicative scaling factor $s$, as defined in Algorithm \ref{alg:multip}. Moderate perturbations improved predictive accuracy, with the lowest MAE achieved at $s = \pm50$. Excessive scaling led to performance degradation due to over-distortion. The best-performing setting (Figure~\ref{fig:multiplicative-scaling-best}) resulted in an 8.0\% reduction in MAE, indicating improved robustness to intensity-based variations.

\begin{figure}[!htbp]
\centering
\begin{minipage}{0.48\textwidth}
\includegraphics[width=\textwidth]{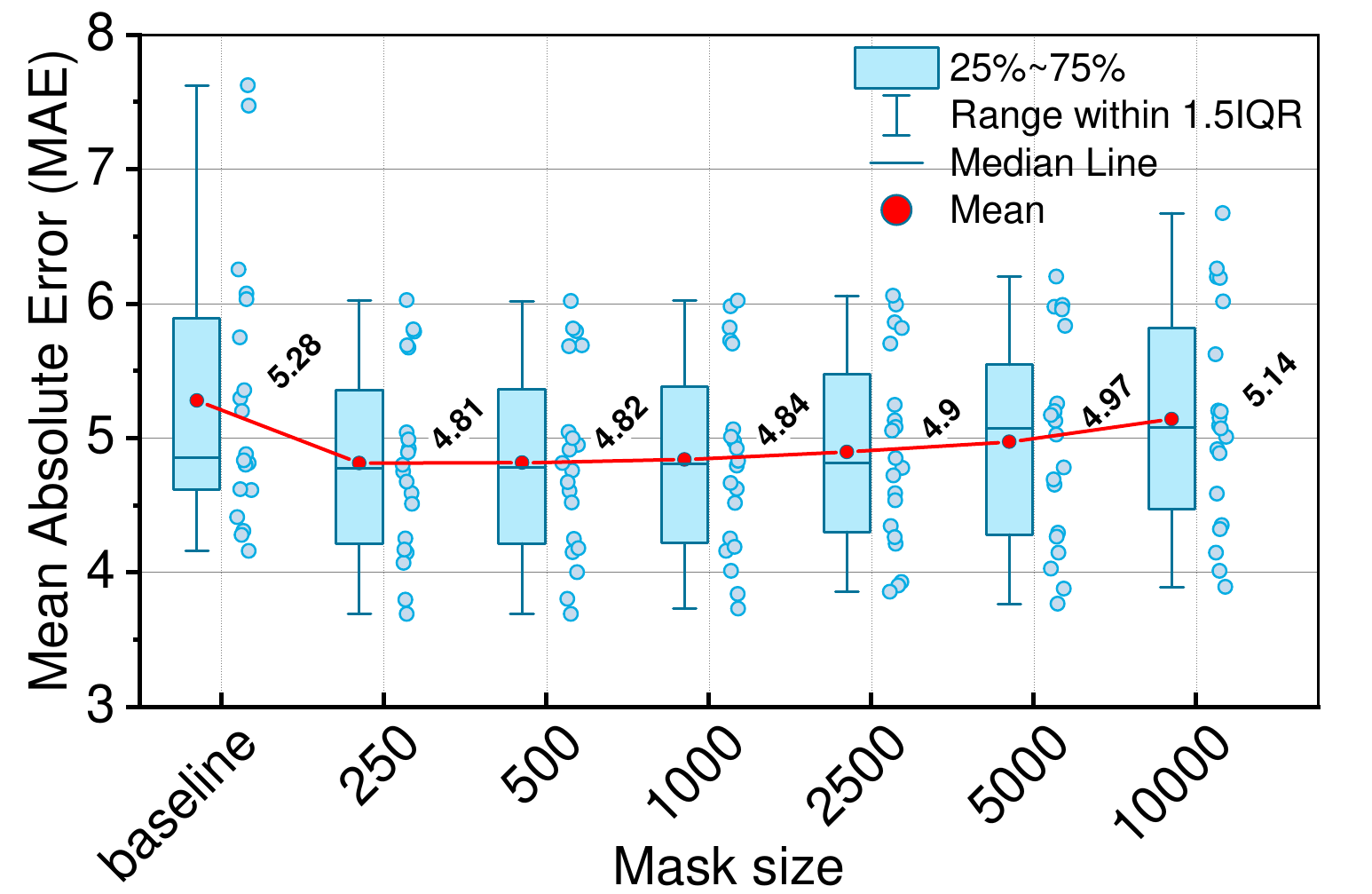}
\caption{Effect of varying the mask size $m$ for masking augmentation on out-of-distribution performance. The boxplot compares the out-of-distribution MAE across 20 runs, each with different random seeds.
}
\label{fig:masking-multiplier}
\end{minipage}
\hfill
\begin{minipage}{0.48\textwidth}
\includegraphics[width=\textwidth]{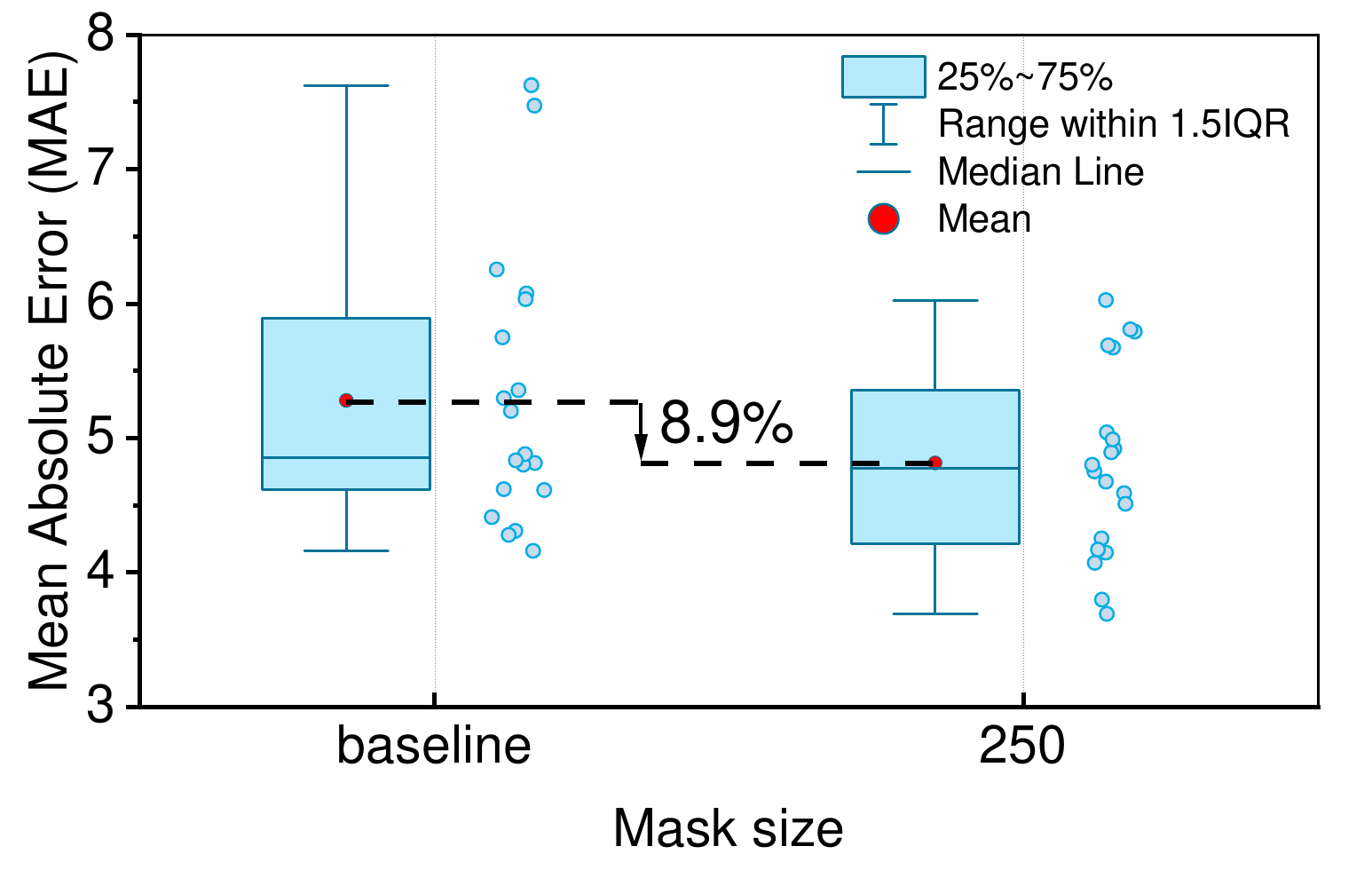}
\caption{Out-of-distribution performance of the best-performing masking augmentation setting ($m$ = 250) compared to the baseline across 20 runs, each with different random seeds. \newline}
\label{fig:masking-best}
\end{minipage}
\end{figure}

\textbf{Masking:} Figure~\ref{fig:masking-multiplier} presents the effect of varying the mask size $m$, as defined in Algorithm~\ref{alg:masking}. The results indicate that masking augmentation improves predictive accuracy at lower mask sizes, with MAE reaching its minimum at $m = 250$. Beyond this value, performance begins to degrade as excessive masking removes too much spectral information, leading to a loss of critical predictive features. The optimal configuration (Figure~\ref{fig:masking-best}) led to an 8.9\% reduction in MAE.

\begin{figure}[!htbp]
\centering
\begin{minipage}{0.48\textwidth}
\includegraphics[width=\textwidth]{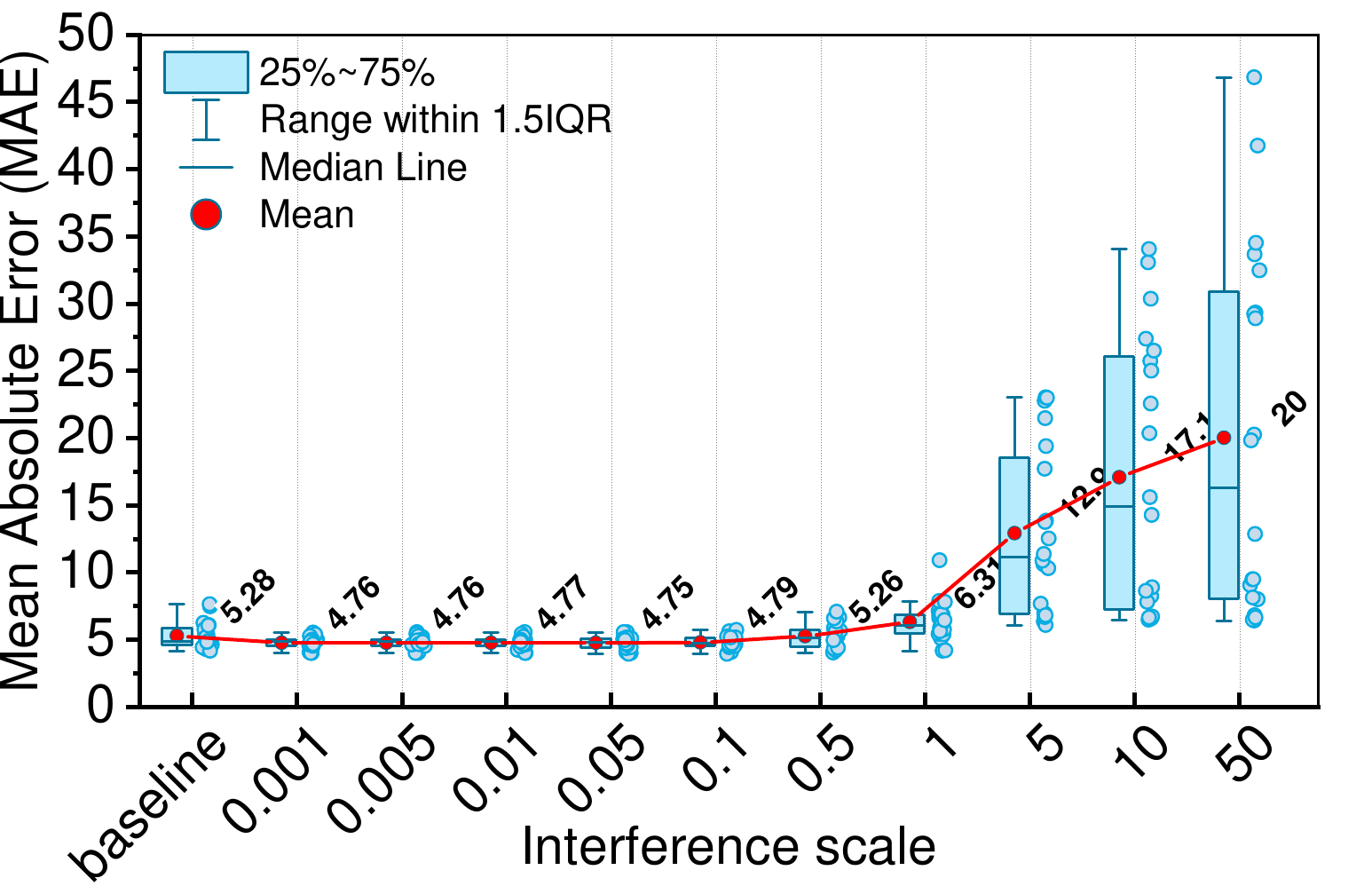}
\caption{Effect of varying the interference scale $c$ for spectral interference augmentation on out-of-distribution performance. The boxplot compares the out-of-distribution MAE across 20 runs, each with different random seeds.}
\label{fig:interference}
\end{minipage}
\hfill
\begin{minipage}{0.48\textwidth}
\includegraphics[width=\textwidth]{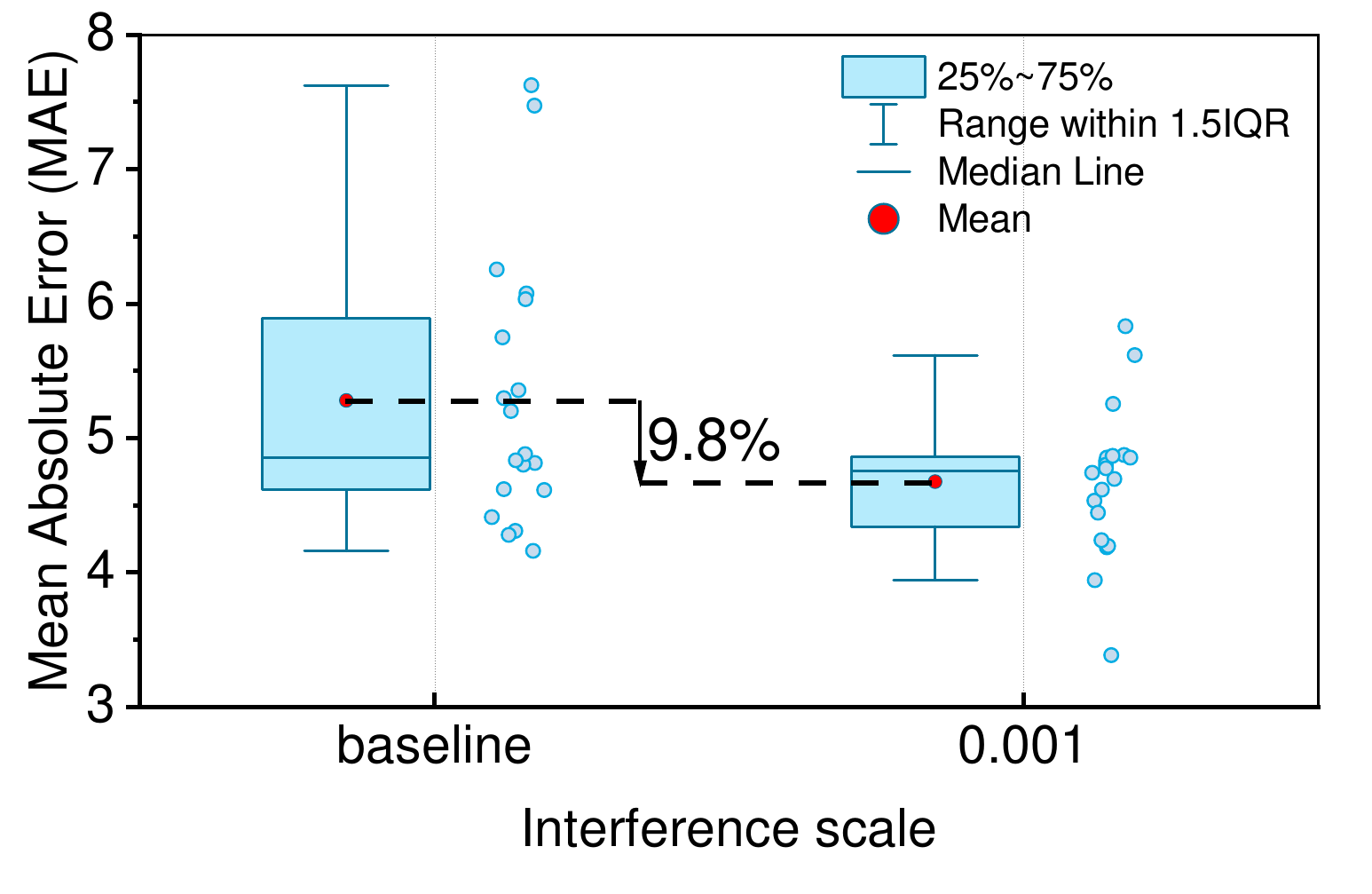}
\caption{Out-of-distribution performance of the best-performing spectral interference augmentation setting ($c$ = 0.001) compared to the baseline across 20 runs, each with different random seeds.\newline}
\label{fig:interference-best}
\end{minipage}
\end{figure}

\textbf{Spectral interference:} The effect of varying the interference scale $c$, as defined in Algorithm~\ref{alg:interference}, is shown in Figure~\ref{fig:interference}. Small values of $c$ (e.g., $0.001$) improved generalization, while larger values degraded performance as larger interference scales introduce excessive noise, corrupting meaningful spectral features. The optimal setting (Figure~\ref{fig:interference-best}) reduced MAE by 9.8\%.

\begin{figure}[!htbp]
\centering
\begin{minipage}{0.48\textwidth}
\includegraphics[width=\textwidth]{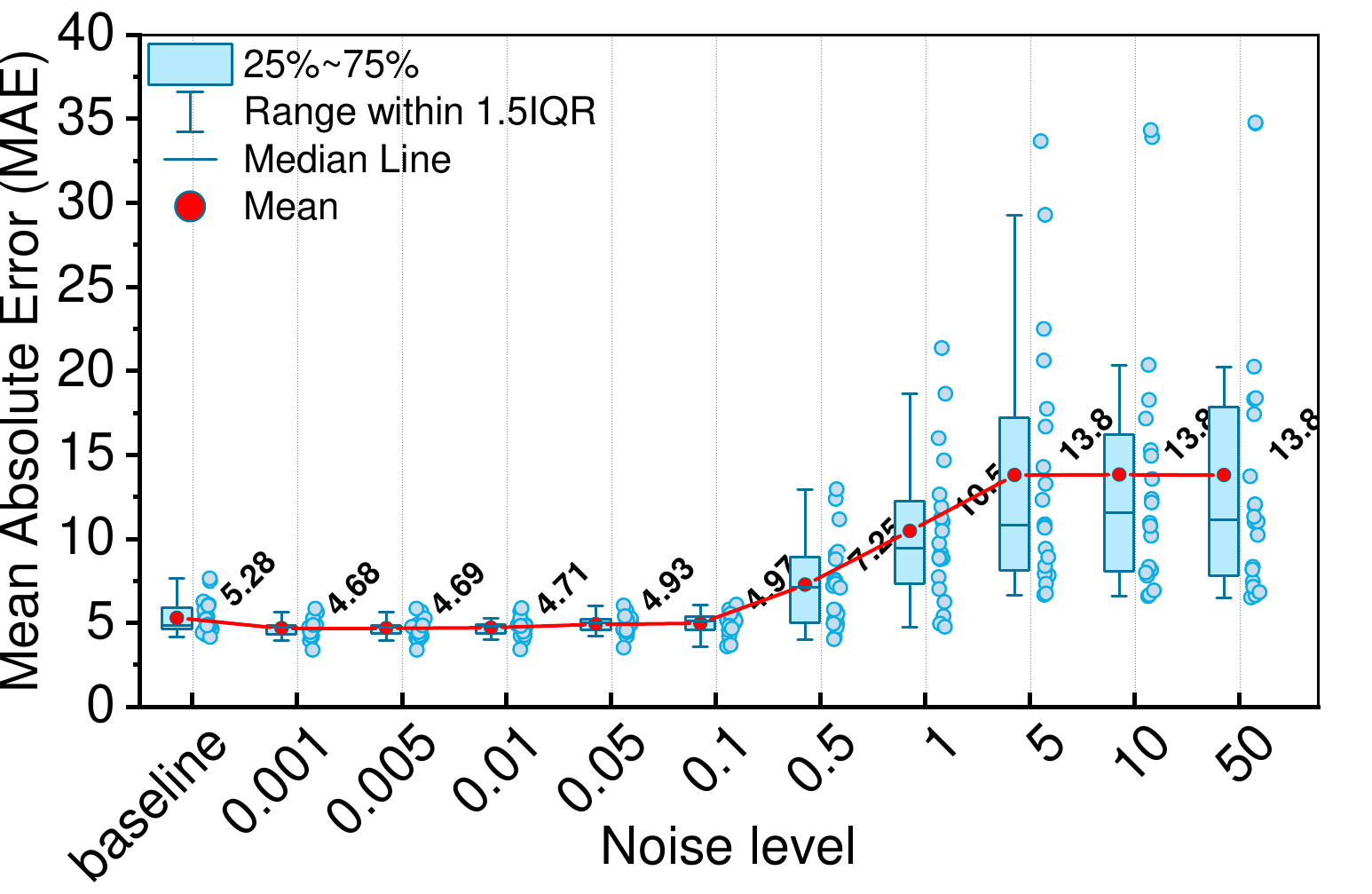}
\caption{Effect of varying the noise level $n$ for additive noise augmentation on out-of-distribution performance. The boxplot compares the out-of-distribution MAE across 20 runs, each with different random seeds.
}
\label{fig:noise-multiplier}
\end{minipage}
\hfill
\begin{minipage}{0.48\textwidth}
\includegraphics[width=\textwidth]{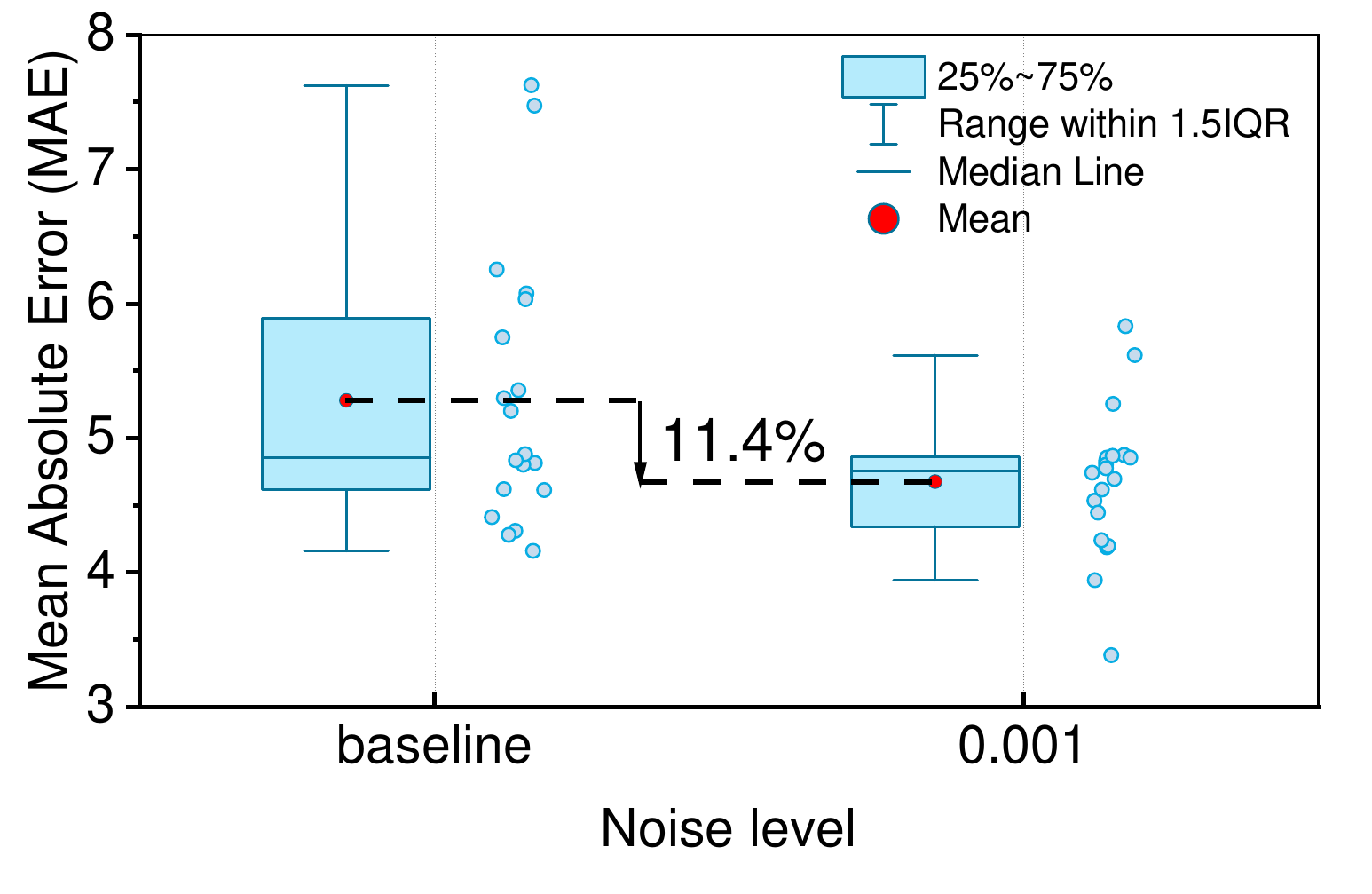}
\caption{Out-of-distribution performance of the best-performing additive noise augmentation setting ($n$ = 0.001) compared to the baseline across 20 runs, each with different random seeds.\newline}
\label{fig:noise-best}
\end{minipage}
\end{figure}

\textbf{Additive noise:} Figure \ref{fig:noise-multiplier} presents the effect of varying $n$, as defined in Algorithm~\ref{alg:noise}, demonstrating that low noise levels enhance predictive accuracy, while excessive noise degrades performance. The model achieves the lowest MAE at $n$ = 0.001, beyond which performance deteriorates as larger noise levels introduce excessive spectral distortions, reducing the model’s ability to extract meaningful features. The best setting (Figure~\ref{fig:noise-best}) led to an 11.4\% reduction in MAE, suggesting greater robustness to noise.

\begin{figure}[!htbp]
\centering
\begin{minipage}{0.48\textwidth}
\includegraphics[width=\textwidth]{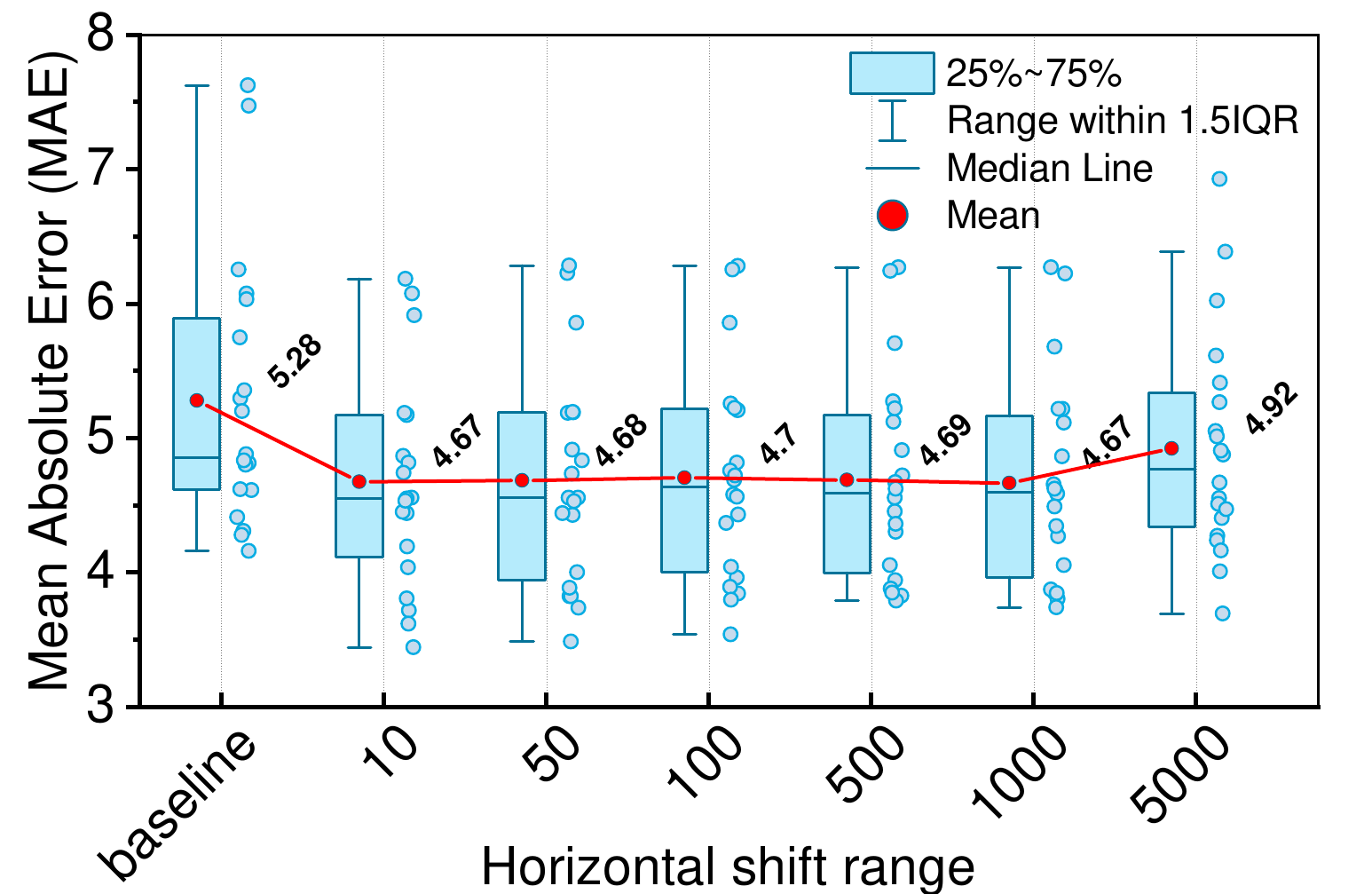}
\caption{Effect of varying the horizontal shift range $h$ for horizontal shift augmentation on out-of-distribution performance. The boxplot compares the out-of-distribution MAE across 20 runs, each with different random seeds.
}
\label{fig:hshift}
\end{minipage}
\hfill
\begin{minipage}{0.48\textwidth}
\includegraphics[width=\textwidth]{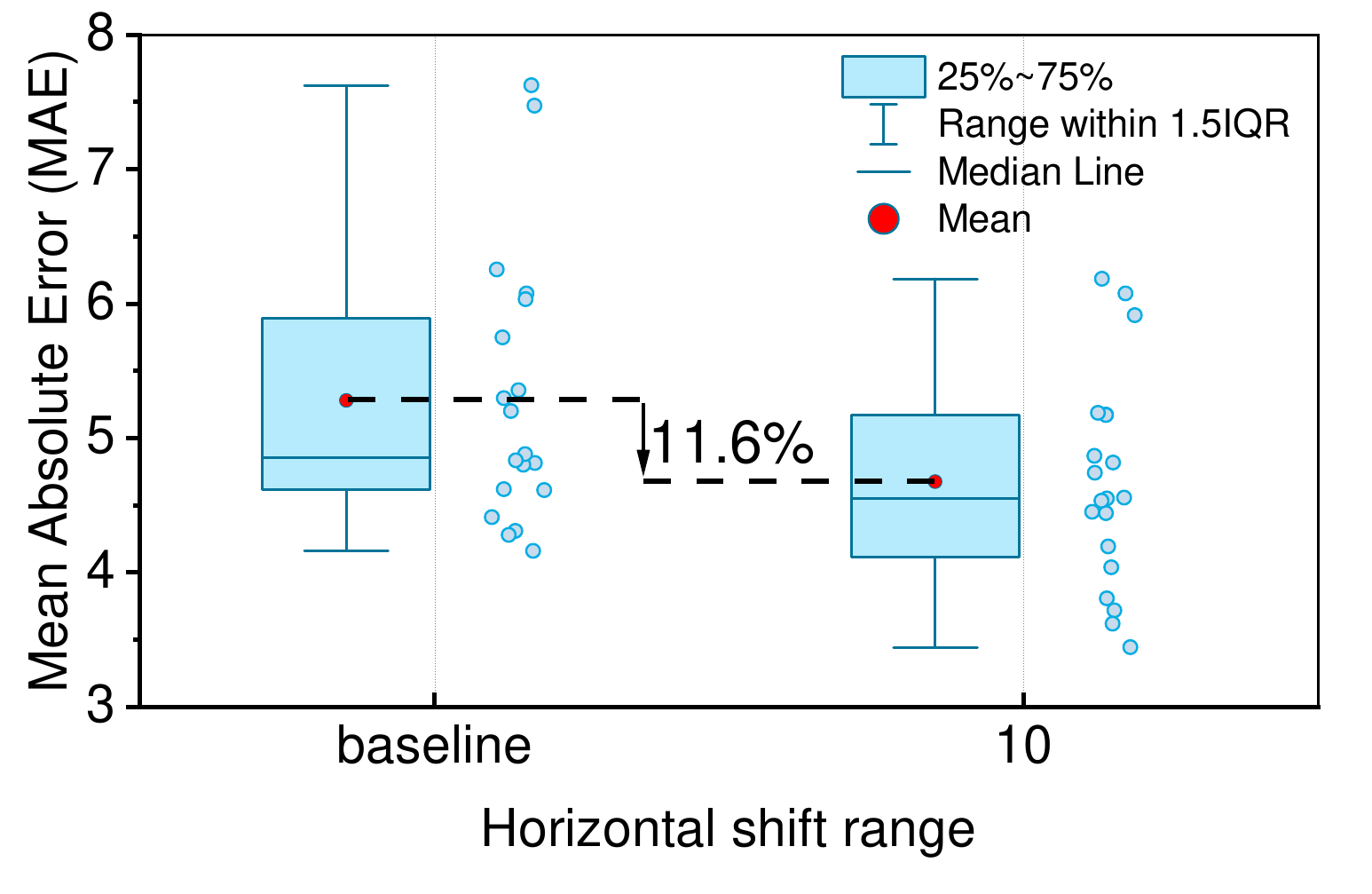}
\caption{Out-of-distribution performance of the best-performing horizontal shift augmentation setting ($h$ = 10) compared to the baseline across 20 runs, each with different random seeds.\newline}
\label{fig:hshift-best}
\end{minipage}
\end{figure}

\textbf{Horizontal shifting:} Figure~\ref{fig:hshift} shows the effect of varying the horizontal shift range $h$, as defined in Algorithm~\ref{alg:h-shift}. A shift of $h = 10$ produced the best result, with further increases yielding minimal gains or slight degradation. The optimal configuration (Figure~\ref{fig:hshift-best}) reduced MAE by 11.6\%.

\begin{figure}[!htbp]
\centering
\begin{minipage}{0.48\textwidth}
\includegraphics[width=\textwidth]{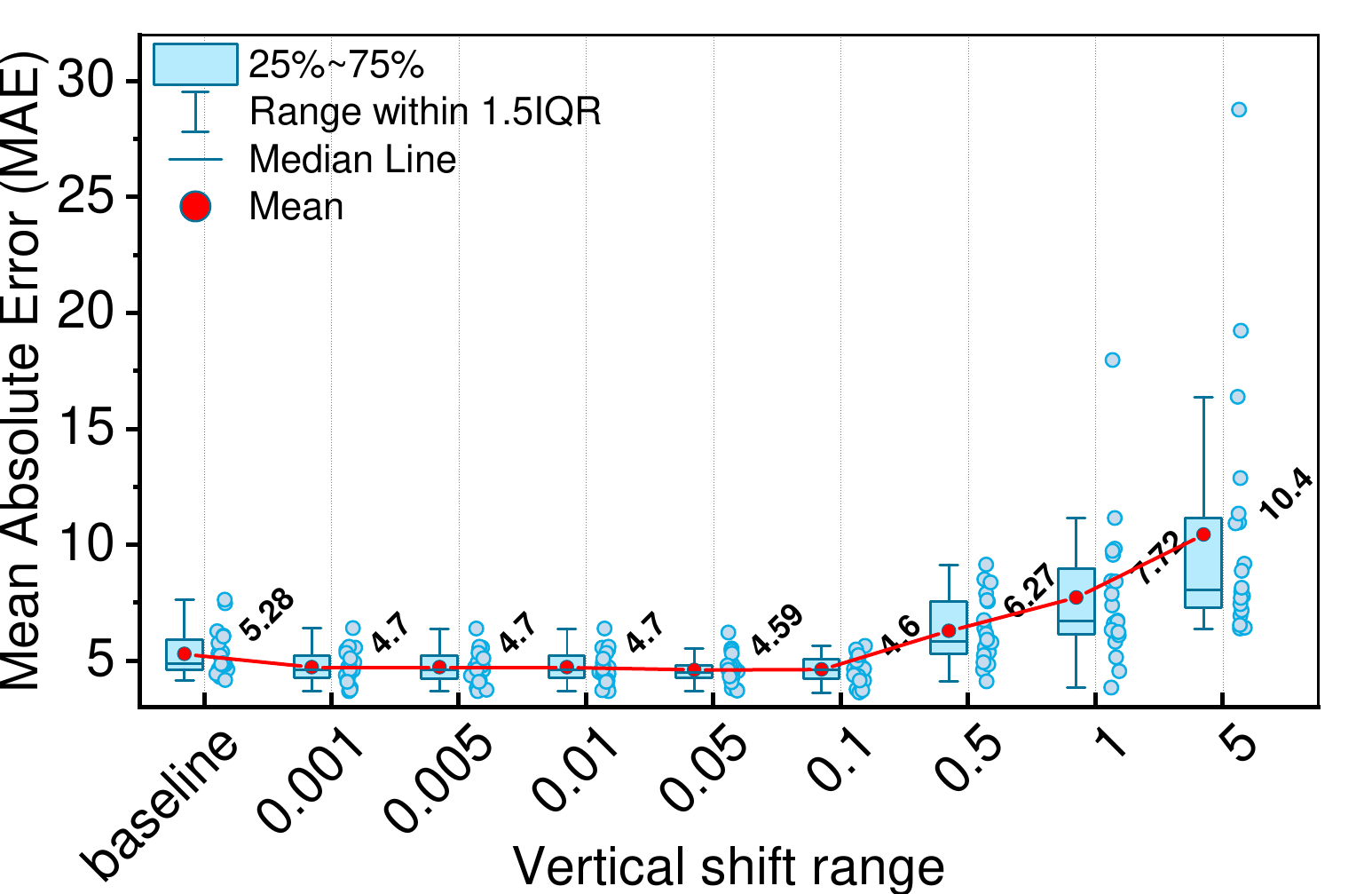}
\caption{Effect of varying the vertical shift range $v$ for vertical shift augmentation on out-of-distribution performance. The boxplot compares the out-of-distribution MAE across 20 runs, each with different random seeds.}
\label{fig:vshift}
\end{minipage}
\hfill
\begin{minipage}{0.48\textwidth}
\includegraphics[width=\textwidth]{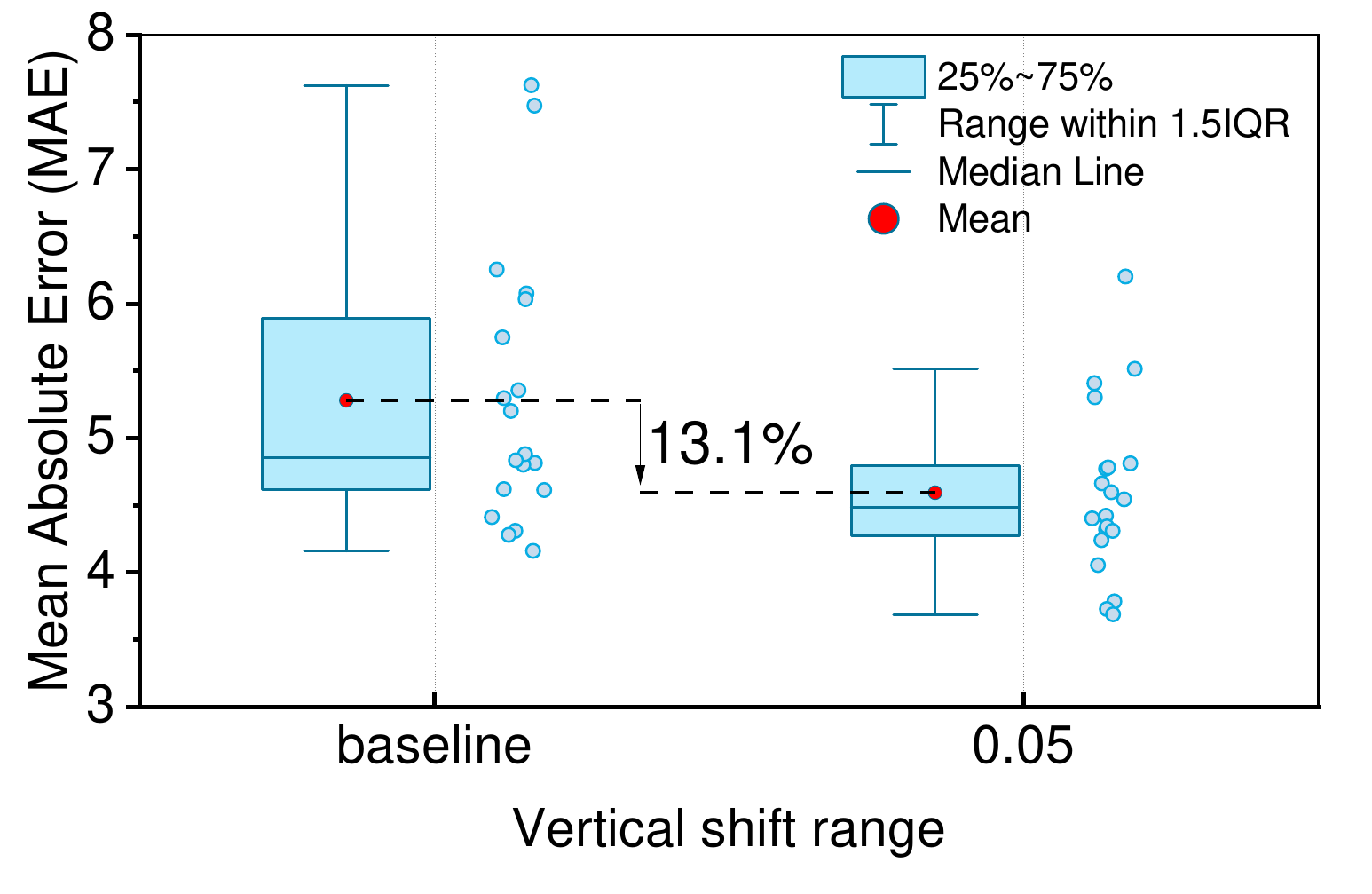}
\caption{Out-of-distribution performance of the best-performing vertical shift augmentation setting ($v$ = 0.05) compared to the baseline across 20 runs, each with different random seeds.\newline}
\label{fig:vshift-best}
\end{minipage}
\end{figure}

\textbf{Vertical shifting:} the vertical shift range $v$, as defined in Algorithm~\ref{alg:v-shift}, was varied (Figure~\ref{fig:vshift}). The best performance was observed at $v = 0.05$, beyond which prediction accuracy declined. The optimal setting (Figure~\ref{fig:vshift-best}) achieved a 13.1\% reduction in MAE, indicating improved robustness to baseline absorbance variations.

\begin{figure}[!htbp]
\centering
\begin{minipage}{0.48\textwidth}
\includegraphics[width=\textwidth]{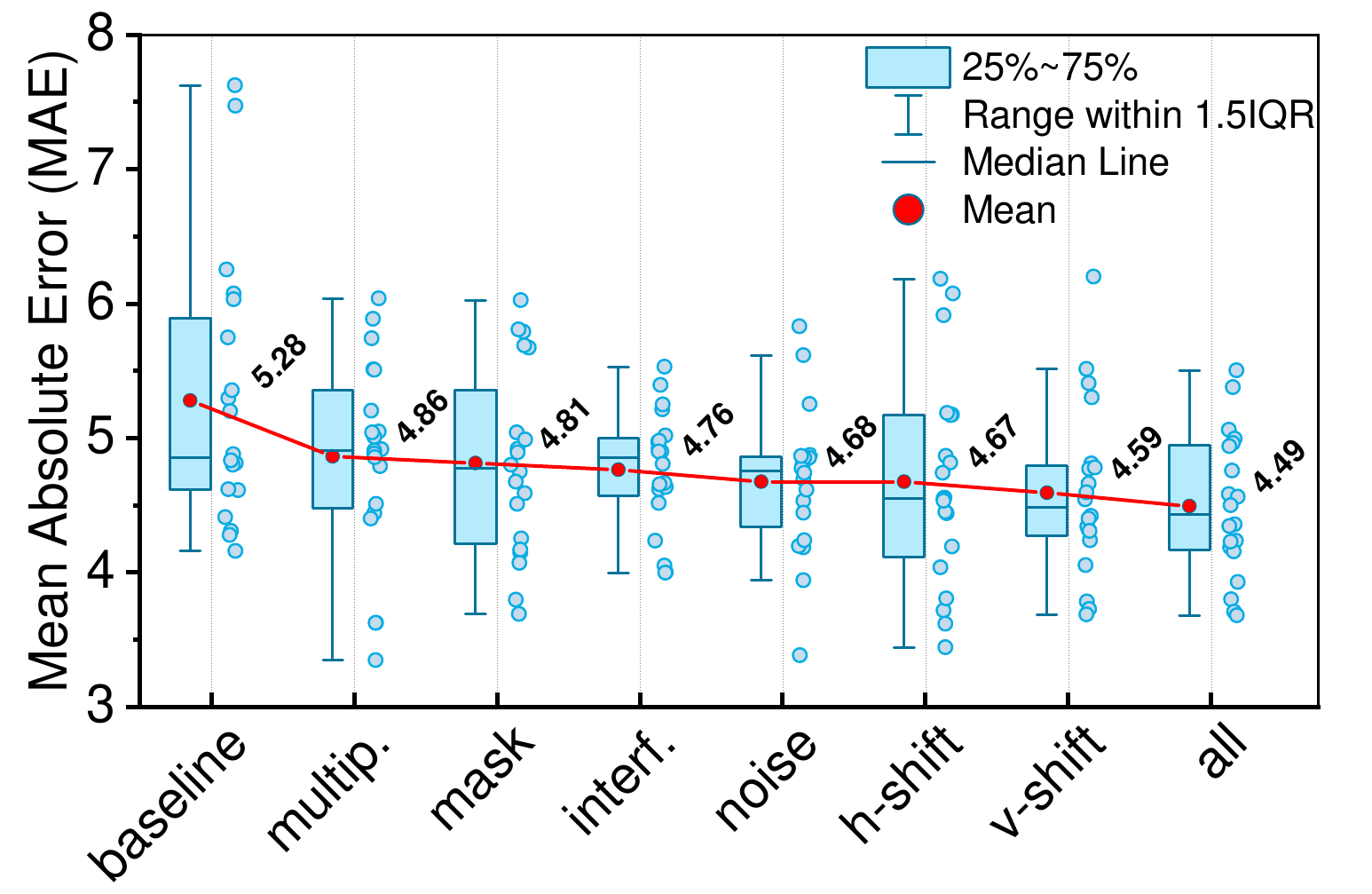}
\caption{Comparison of baseline performance, models trained with individual augmentation types, and a model trained with all augmentations combined across 20 runs, each with different random seeds.}
\label{fig:all-augmentations}
\end{minipage}
\hfill
\begin{minipage}{0.48\textwidth}
\includegraphics[width=\textwidth]{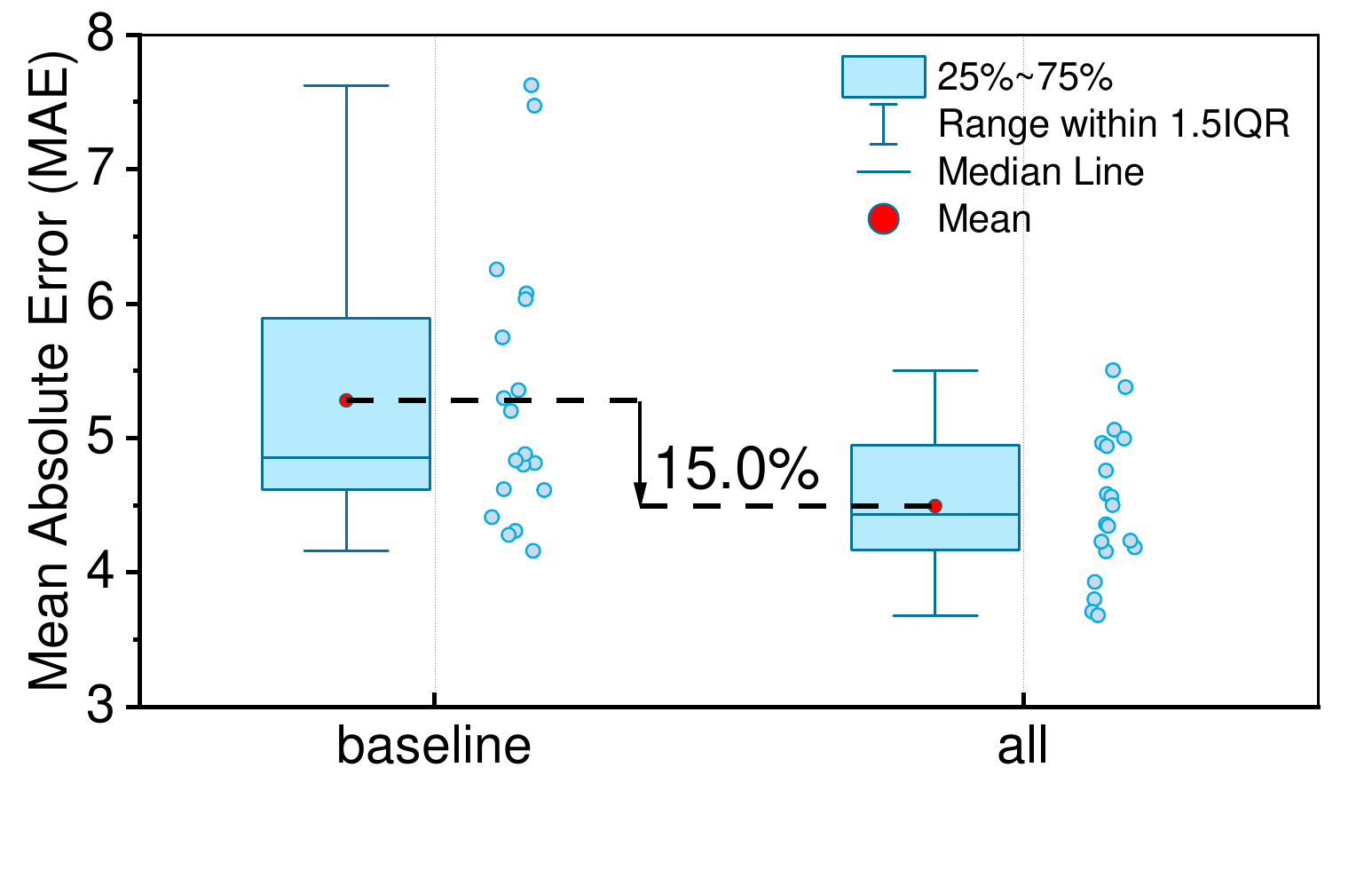}
\caption{Out-of-distribution performance of the model trained with all augmentations combined, compared to the baseline across 20 runs, each with different random seeds.\newline}
\label{fig:all-augmentations-best}
\end{minipage}
\end{figure}

To conclude the discussion on unsupervised data augmentation, the effect of combining all augmentation types in a single model was evaluated to determine whether simultaneous application of multiple strategies could improve predictive accuracy beyond the gains observed from individual augmentations. Figure \ref{fig:all-augmentations} compares the baseline model, models trained with each augmentation type individually, and a model incorporating all augmentations simultaneously. All individual augmentations contributed to MAE reduction, with vertical and horizontal shifts yielding the largest improvements. However, the combined augmentation model achieved the lowest overall MAE, with a 15.0\% reduction compared to the baseline, outperforming any single augmentation, as shown in Figure \ref{fig:all-augmentations-best}. This significant improvement suggests that enforcing consistency across multiple perturbations allows the model to learn more invariant and generalizable spectral features.

\subsubsection{Fuelprop}

The final evaluation assesses the composite model, referred to as Fuelprop, which integrates all four data enhancement strategies: synthetic spectra blending, pseudo-labeling, synthetic data generation, and unsupervised data augmentation. The objective of Fuelprop is to leverage the strengths of each individual strategy to maximize predictive accuracy and generalization to out-of-distribution samples.

\begin{figure}[!htbp]
\centering
\begin{minipage}{0.48\textwidth}
\includegraphics[width=\textwidth]{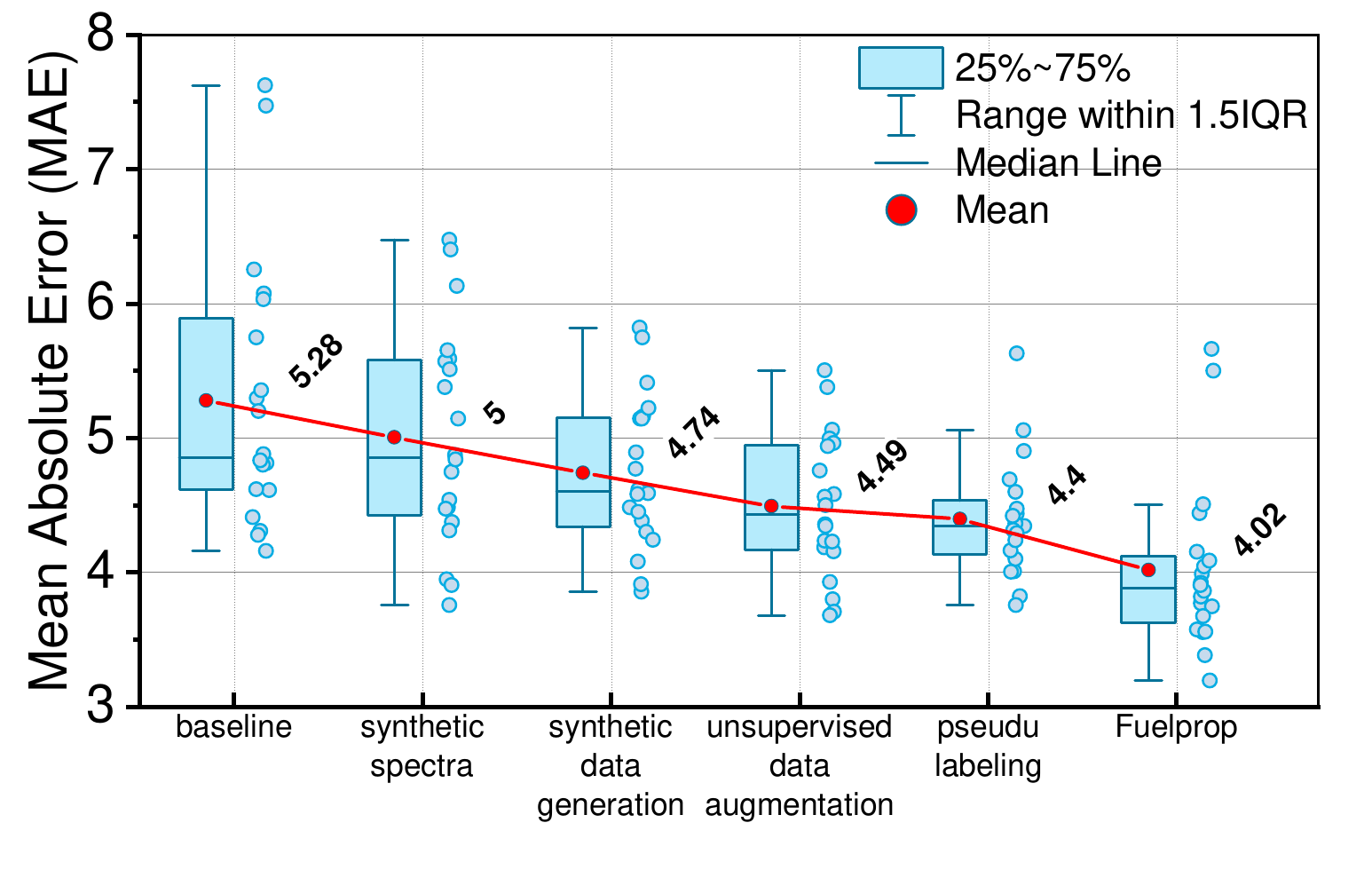}
\caption{Comparison of the baseline model, models trained with individual enhancement strategies, and Fuelprop. The results illustrate the cumulative performance gains achieved by integrating multiple data enhancement techniques.}
\label{fig:fuelprop-comparison}
\end{minipage}
\hfill 
\begin{minipage}{0.48\textwidth}
\includegraphics[width=\textwidth]{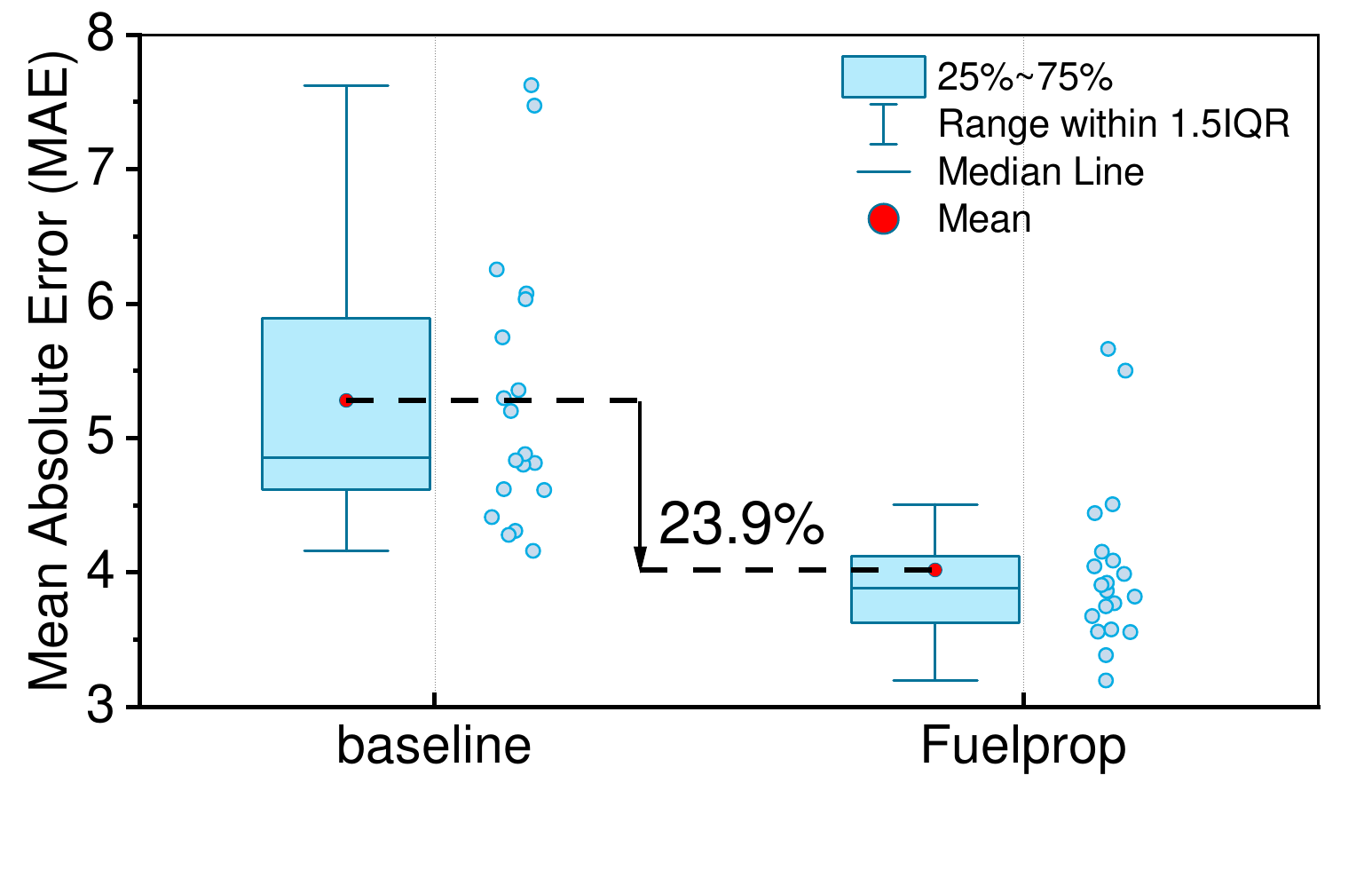}
\caption{Direct comparison of the baseline model and Fuelprop across 20 runs, each with different random seeds.
\newline
\newline
\newline
}
\label{fig:fuelprop}
\end{minipage}
\end{figure}

Figure \ref{fig:fuelprop-comparison} presents a comparison of the baseline model and models trained with each individual enhancement strategy. The results indicate an improvement in performance as each strategy is incorporated, with Fuelprop achieving the lowest MAE showing a 23.9\% reduction relative to the baseline as shown in Figure \ref{fig:fuelprop}. The observed reductions in MAE suggest that each strategy contributes unique benefits to model performance, and their combination provides a compounded effect.

\begin{figure}[!htbp] 
\centering 
\includegraphics[width=0.6\textwidth]{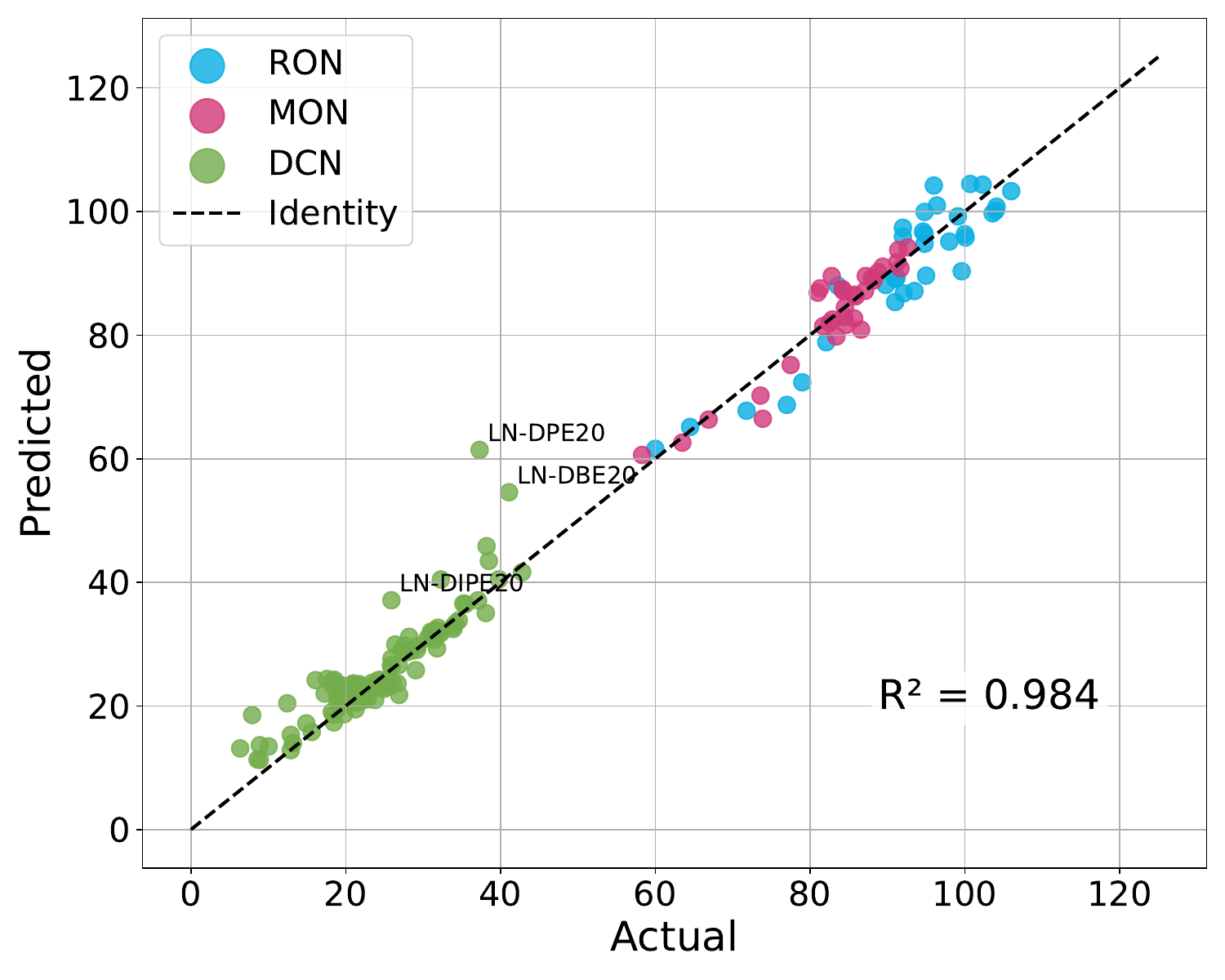} \caption{Actual vs. predicted values of RON, MON, and DCN for the Fuelprop model, evaluated on real fuels held out from training. Points with residuals exceeding $\pm10$ are labeled and correspond to blends of light naphtha J80 with dipropyl ether, dibutyl ether, and diisopropyl ether.} 
\label{fig:fuelprop-parity} 
\end{figure}

Further analysis of Fuelprop’s predictive capabilities is shown in Figure \ref{fig:fuelprop-parity}, which presents a comparison of predicted versus actual values for RON, MON, and DCN. The model achieves a high level of accuracy, with an $R^2$ value of 0.984. Data points labeled in the plot represent predictions with residuals exceeding ± 10. These outliers are:

\begin{itemize}
\item LN-DPE20: A blend of 80\% light naphtha J80 and 20\% dipropyl ether (DPE) by volume.
\item LN-DBE20: A blend of 80\% light naphtha J80 and 20\% dibutyl ether (DBE) by volume.
\item LN-DIPE20: A blend of 80\% light naphtha J80 and 20\% diisopropyl ether (DIPE) by volume.
\end{itemize}
The model exhibited larger prediction errors for these specific blends. Notably, the DCN value of DPE was not included in the training dataset, which could contribute to the observed deviation in predictions for LN-DPE20.

\section{Conclusion}

This study presents a comprehensive investigation into the prediction of fuel properties (RON, MON, and DCN) using ATR-FTIR spectroscopy and advanced machine learning techniques. We have addressed the critical need for high-quality, diverse datasets for training robust predictive models by curating a substantial ATR-FTIR spectral dataset that includes pure components, surrogate blends, and real fuels and making it publicly available. Our analysis highlights the significant impact of data enhancement strategies on model performance. Specifically, the Fuelprop model, which combines data imputation through synthetic spectra blending and pseudo-labeling, semi-supervised synthetic data generation, and consistency enforcement by unsupervised data augmentation, achieved a 23.9\% improvement in prediction accuracy compared to a baseline CNN model trained without any of these strategies. Individual data enhancement strategies also showed substantial improvements, ranging from 5.3\% to 16.7\%, showcasing their individual effectiveness in mitigating data limitations. These results emphasize the potential of ATR-FTIR spectroscopy, coupled with effective data enhancement and advanced machine learning, for accurate and efficient fuel property prediction.

The Fuelprop model and the developed dataset provide a significant step forward in the advancement chemometric methods for fuel property prediction, which has the potential to streamline synthetic fuel development and reduce reliance on expensive and time-consuming conventional testing methods. Future work will focus on refining these models to improve their accuracy and generalization across a broader range of fuel types, allowing for wider adoption of these efficient characterization techniques.

\textbf{Acknowledgement}
This work was funded by the Office of Sponsored Research at King Abdullah University of Science and Technology (KAUST).

\bibliographystyle{elsarticle-num-names} 
\bibliography{ref}

\end{document}